\newcommand{\sersic}{S\'ersic\ }

\documentclass[longauth]{aa}  

\usepackage{graphicx}
\usepackage{txfonts}
\usepackage[colorlinks,citecolor=blue,hypertexnames=true,bookmarks=false,breaklinks=true]{hyperref}
\begin{document} 

   \title{CEERS: MIRI deciphers the spatial distribution of dust-obscured star formation in galaxies at $0.1<z<2.5$}

   \author{Benjamin Magnelli\inst{1}
          \and
          Carlos G\'omez-Guijarro\inst{1}
          \and
          David Elbaz\inst{1}
          \and
          Emanuele Daddi\inst{1}
          \and
          Casey Papovich\inst{2,3}
          \and
          Lu Shen\inst{2,3}
          \and
          Pablo Arrabal Haro\inst{4}
          \and
          Micaela B. Bagley\inst{5}
          \and
          Eric F. Bell\inst{6}
          \and
          V\'eronique Buat\inst{7}
          \and
          Luca Costantin\inst{8}
          \and
          Mark Dickinson\inst{4}
          \and
          Steven L. Finkelstein\inst{5}
          \and
          Jonathan P. Gardner\inst{9}
          \and
          Eric F. Jim\'enez-Andrade\inst{10}
          \and
          Jeyhan S. Kartaltepe\inst{11}
          \and
          Anton M. Koekemoer\inst{12}
          \and
          Yipeng Lyu\inst{1}
          \and
          Pablo G. P\'erez-Gonz\'alez\inst{8}
          \and
          Nor Pirzkal\inst{13}
          \and
          Sandro Tacchella\inst{14,15}
          \and
          Alexander de la Vega\inst{16}
          \and
          Stijn Wuyts\inst{17}
          \and
          Guang Yang\inst{18,19}
          \and
          L. Y. Aaron Yung\inst{9\thanks{NASA Postdoctoral Fellow}}
          \and
          Jorge Zavala\inst{20}
          }

   \institute{Universit{\'e} Paris-Saclay, Universit{\'e} Paris Cit{\'e}, CEA, CNRS, AIM, 91191, Gif-sur-Yvette, France\\
    \email{benjamin.magnelli@cea.fr}
    \and
    Department of Physics and Astronomy, Texas A\&M University, College Station, TX, 77843-4242 USA
    \and
    George P. and Cynthia Woods Mitchell Institute for Fundamental Physics and Astronomy, Texas A\&M University, College Station, TX, 77843-4242 USA
    \and
    NSF's National Optical-Infrared Astronomy Research Laboratory, 950 N. Cherry Ave., Tucson, AZ 85719, USA
    \and
    Department of Astronomy, The University of Texas at Austin, Austin, TX, USA
    \and
    Department of Astronomy, University of Michigan, 1085 South University Avenue, Ann Arbor, MI 48109-1107, USA
    \and
    Aix Marseille Univ, CNRS, CNES, LAM Marseille, France
    \and
    Centro de Astrobiolog\'ia (CAB), CSIC-INTA, Ctra. de Ajalvir km 4, Torrej\'on de Ardoz, E-28850, Madrid, Spain
    \and
    Astrophysics Science Division, NASA Goddard Space Flight Center, 8800 Greenbelt Rd, Greenbelt, MD 20771, USA
    \and
    Instituto de Radioastronom\'ia y Astrof\'isica, Universidad Nacional Aut\'onoma de M\'exico, Antigua Carretera a P\'atzcuaro \#8701, Ex-Hda. San Jos\'e de la Huerta, Morelia, Michoac\'an, M\'exico C.P. 58089
    \and
    Laboratory for Multiwavelength Astrophysics, School of Physics and Astronomy, Rochester Institute of Technology, 84 Lomb Memorial Drive, Rochester, NY 14623, USA
    \and
    Space Telescope Science Institute, 3700 San Martin Dr., Baltimore, MD 21218, USA
    \and
    ESA/AURA Space Telescope Science Institute
    \and
    Kavli Institute for Cosmology, University of Cambridge, Madingley Road, Cambridge, CB3 0HA, UK
    \and
    Cavendish Laboratory, University of Cambridge, 19 JJ Thomson Avenue, Cambridge, CB3 0HE, UK
    \and
    Department of Physics and Astronomy, University of California, 900 University Ave, Riverside, CA 92521, USA
    \and
    Department of Physics, University of Bath, Claverton Down, Bath, BA2 7AY, UK
    \and
    Kapteyn Astronomical Institute, University of Groningen, P.O. Box 800, 9700 AV Groningen, The Netherlands
    \and
    SRON Netherlands Institute for Space Research, Postbus 800, 9700 AV Groningen, The Netherlands
    \and
    National Astronomical Observatory of Japan, 2-21-1 Osawa, Mitaka, Tokyo 181-8588, Japan
}

\date{Received ; accepted}

 
  \abstract
   {}
   {We study the stellar (i.e., rest-optical) and dust-obscured star-forming (i.e., rest-mid-infrared) morphologies (i.e., sizes and \sersic indices) of star-forming galaxies (SFGs) at $0.1<z<2.5$.}
   {We combined \textit{Hubble Space Telescope} (\textit{HST}) images from the Cosmic Assembly Near-infrared Deep Extragalactic Legacy Survey (CANDELS) with \textit{JWST} images from the Cosmic Evolution Early Release Science (CEERS) survey to measure the stellar and dust-obscured star formation distributions of 69 SFGs.
   Rest-mid-infrared (rest-MIR) morphologies were determined using a Markov chain Monte Carlo (MCMC) approach applied to the sharpest Mid-InfraRed Instrument (MIRI) images (i.e., shortest wavelength) dominated by dust emission ($S_{\nu}^{\rm dust}/S_{\nu}^{\rm total} >75\%$), as inferred for each galaxy from our optical-to-far-infrared spectral energy distribution fits with \texttt{CIGALE}. 
   Rest-MIR \sersic indices were only measured for the brightest MIRI sources, that is, with a signal-to-noise ($S/N$) greater than 75 (35 galaxies).
   At a lower $S/N$, simulations do indeed show that simultaneous measurements of both the size and \sersic index become less reliable. 
   We extended our study to fainter sources (i.e., $S/N>10$; 69 galaxies) by restricting our structural analysis to their rest-MIR sizes ($Re_{\rm MIR}$) and by fixing their \sersic index to a value of one.}
   {Our MIRI-selected sample corresponds to a mass-complete sample ($>80\%$) of SFGs down to stellar masses $10^{9.5}, 10^{9.5},$ and $10^{10}\,M_{\odot}$ at $z\sim0.3$, 1, and 2, respectively. 
   The rest-MIR \sersic index of bright galaxies ($S/N>75$) has a median value of $0.7_{-0.3}^{+0.8}$ (the range corresponds to the 16th and 84th percentiles), which is in good agreement with their median rest-optical \sersic indices.
   The \sersic indices as well as the distribution of the axis ratio of these galaxies suggest that they have a disk-like morphology in the rest-MIR.
   Galaxies above the main sequence (MS) of star formation (i.e., starbursts) have rest-MIR sizes that are, on average, a factor $\sim\,$2 smaller than their rest-optical sizes ($Re_{\rm Opt.}$).
   The median rest-optical to rest-MIR size ratio of MS galaxies increases with their stellar mass, from $1.1_{-0.2}^{+0.4}$ at $\sim10^{9.8}\,M_{\odot}$ to $1.6_{-0.3}^{+1.0}$ at $\sim10^{11}\,M_{\odot}$.
   This mass-dependent trend resembles the one found in the literature between the rest-optical and rest-near-infrared sizes of SFGs, suggesting that it is primarily due to radial color gradients affecting rest-optical sizes and that the sizes of the stellar and star-forming components of SFGs are, on average, consistent at all masses.
   There is, however, a small population of SFGs ($\sim\,$$15\%$) with a compact star-forming component embedded in a larger stellar structure, with $Re^{\rm c}_{\rm Opt.}$$\,>\,$$1.8$$\,\times\,$$Re_{\rm MIR}$. 
   This population could be the missing link between galaxies with an extended stellar component and those with a compact stellar component, the so-called blue nuggets.}
   {}

   \titlerunning{Rest-mid-infrared sizes of star-forming galaxies since $z\sim2.5$}
   
   \keywords{ galaxies: evolution -- galaxies: high-redshift -- galaxies: structure -- infrared: galaxies}

   \maketitle
%

\section{Introduction}
In the star formation rate (SFR)--stellar mass ($M_\ast$) plane, galaxies roughly divide into two groups, star-forming galaxies (SFGs) and quiescent galaxies (QGs).
SFGs are mostly associated with disk-dominated systems \citep[i.e., with \sersic indices $n\sim1$; e.g.,][]{Wuyts.2011} and they lie on a tight locus of the SFR--$M_\ast$ plane, known as the main sequence (MS) of SFGs \citep[e.g.,][and references therein]{Noeske.2007,Elbaz.2007,Schreiber.2015,Barro.2019}. 
The small scatter of the MS, observed out to $z\sim4$ \citep{Schreiber.2015}, suggests that secular evolution is the dominant mode of growth in SFGs, where gas inflow, outflow, and the star formation are in equilibrium \citep[e.g.,][]{Bouche.2010,Dave.2012,Lilly.2013,Peng.2014,Rathaus.2016}. 
Conversely, QGs are mostly associated with massive bulge-dominated systems ($n\gtrsim3$) with no or little star formation and they lie well below the MS \citep[e.g.,][]{Wuyts.2011}.
The rest-optical sizes of both SFGs and QGs show a tight but distinct scaling with galaxy stellar mass \citep[e.g.,][]{Wel.2014}.
At all masses, SFGs are larger than QGs, and their size--mass relation has a shallower slope and its intercept has a slower redshift evolution than those of QGs \citep[e.g.,][]{Wel.2014}.
Connecting progenitors and descendants implies that individual SFGs must have experienced a significant size growth \citep{Dokkum.2013,Dokkum.2015} while evolving along the MS, and that, on their way to quiescence, SFGs must experience a phase of significant compaction \citep[e.g.,][]{Cheung.2012,Barro.2017,Mosleh.2017}.
The mechanisms at the origin of this otherwise well-established structural evolution of SFGs are far from being clearly understood.

To investigate the processes leading to the structural evolution of SFGs, it is necessary to measure not only their current morphologies (i.e., those of their stellar components), but also the distribution of their ongoing star formation, as this allows us to predict their size growth \citep[e.g.,][]{Wilman.2020}.
To this end, tremendous efforts have been made over the past decade to obtain unobscured star formation maps of large samples of high-redshift SFGs from their $H\alpha$ line emission using \textit{Hubble Space Telescope} (\textit{HST}) WFC3 grism spectroscopy or adaptive optics-assisted ground-based integral field unit surveys \citep[e.g.,][]{Nelson.2012,Nelson.2013,Tacchella.2015,Nelson.2016,Tacchella.2018,Wilman.2020,Matharu.2022}.
All of these studies consistently find that the effective radius ($Re$; equivalently half-light radius) of the $H\alpha$ emission is slightly more extended than the stellar continuum, with a weak dependence on mass.
In the absence of dust attenuation, such a profile would translate into a centrally depressed specific SFR (sSFR; i.e., SFR/$M_\ast$) and this would point to an inside-out growth of the stellar disk.
However, SFGs at high redshifts are far from being devoid of dust and are therefore affected by nontrivial radial attenuation gradients \citep{Wuyts.2012, Nelson.2016, Tacchella.2018,Perez-Gonzalez.2023,Zhang.2023}.
By accounting for dust attenuation in these $H\alpha$ measurements, using either the Balmer decrement or the UV slope $\beta$, a converging picture is emerging, in which the sSFR profile of intermediate-mass ($\sim10^{10-11}\,M_\odot$) galaxies is relatively flat--at odds with an inside-out growth scenario-- and in which only massive ($\gtrsim10^{11}\,M_\odot$) galaxies exhibit a centrally depressed sSFR, possibly associated with inside-out quenching \citep{Tacchella.2016,Tacchella.2018}.

Having in mind the effect of dust and the extreme difficulty of correcting for it using simple Balmer decrement or $\beta$-slope prescriptions for massive galaxies \citep[i.e., $>10^{10.5}\,M_\odot$;][]{Wuyts.20110yjd}, other studies have focused their efforts in measuring the dust-obscured star formation distribution of SFGs from their dust emission using the Atacama Large Millimeter Array \citep[ALMA; e.g.,][see also Murphy et al. \citeyear{Murphy.2017}; Jim\'enez-Andrade et al. \citeyear{Jimenez-Andrade.2019} for a radio view on this topic]{Simpson.2015,Ikarashi.2015,Hodge.2016,Fujimoto.2017, Gomez-Guijarro.2018,Elbaz.2018, Lang.2019, Rujopakarn.2019,Puglisi.2019,Franco.2020, Chang.2020, Chen.2020, Tadaki.2020,Puglisi.2021, Gomez-Guijarro.2022}. 
Despite some quantitative differences, all of these studies qualitatively agree on the fact that high-redshift (i.e., $z\gtrsim1$) massive ($M_\ast\gtrsim10^{11}\,M_\odot$) SFGs have dust-obscured star formation distributed in regions which are significantly more compact relative to the size--mass relation of SFGs.
Such compact star-forming core could correspond to the compaction phase that massive SFGs must experience prior to falling into quiescence. 
However, these results draw a very different picture from that which emerges from the $H\alpha$ observations, even after having accounted for dust attenuation.
The reason for such inconsistencies could be twofold: 
(i) $H\alpha$ measurements could still be affected by non-recoverable dust attenuation effects in massive galaxies \citep[e.g.,][]{Tacchella.2022};
and/or (ii) inconsistencies between the rest-(sub)millimeter and $H\alpha$ observations could be due to differences between samples that fit into a common evolutionary sequence.
To make a significant step forward in this domain, we need to probe the dust-obscured star formation distribution of a larger number of SFGs at intermediate mass ($10^{10-11}\,M_\odot$).
Unfortunately, such measurements have been difficult to obtain until now because the detection at high angular resolution of the dust emission from a large sample of SFGs at intermediate mass is observationally very expensive with the very small field of view of ALMA. 

The Mid-InfraRed Instrument (MIRI) on board \textit{JWST} \citep{Gardner.2006} with its unparalleled sensitivity and high angular resolution at mid-infrared (MIR) wavelengths offers a unique opportunity to make significant progress in this domain. 
The MIR emission probed by these observations is indeed dominated by the polycyclic aromatic hydrocarbons (PAH) features of SFGs up to $z\sim2.5$, which are known to provide relatively accurate measurements of the dust-obscured SFR \citep[e.g.,][]{Elbaz.2010, Elbaz.2011, Nordon.2012, Schreiber.2017q5r}.
Furthermore, the angular resolution of, for example, $0\farcs4$ at $15\,\mu$m (3.2\,kpc at $z=1$) should be sufficient to measure the \sersic index (for the brightest galaxies) and effective radius of SFGs at $M_\ast\gtrsim10^{9.5}\,M_\odot$, as these systems have effective radii larger than 3\,kpc at rest-optical wavelength \citep[e.g.,][]{Wel.2014}.
The ability of MIRI to reveal the dust-obscured star formation morphology of high-redshift SFGs was unambiguously demonstrated by \citet[][see also Liu et al.~\citeyear{Liu.2023}]{Shen.2023}.
In their study, \citet{Shen.2023} used preliminary MIRI observations from the Cosmic Evolution Early Release Science (CEERS) survey to successfully measure the dust-obscured star formation distribution of a UV-selected sample of 70 SFGs at $0.2<z<2.5$ and to compare these distributions to the unobscured star formation and stellar distributions.
They find that the effective radii at rest-MIR are slightly smaller ($10\%$) than those at rest-optical at high mass ($\sim10^{10}\,M_\odot$) and find evidence that massive SFGs had an increased fraction of obscured star formation in their inner regions.

In this paper, we extend the analysis of \citet{Shen.2023} to the full set of MIRI observations from CEERS (i.e., four pointings instead of two, covering a total of $\sim8\,$arcmin$^2$) and focus on a mass-complete sample of SFGs instead of a UV-selected sample, which is biased against massive and strongly obscured SFGs.
With this dataset, we measure the rest-MIR morphology of a mass-complete sample of 69 SFGs down to $\sim10^{10}\,M_\odot$ and up to $z\sim2.5$.
At these intermediate masses, the fraction of unobscured star formation is subdominant \citep[$\lesssim30\%$;][and our analysis]{Whitaker.2017,Shen.2023} and, therefore, our rest-MIR morphologies trace, to first order, the total star-forming component of these SFGs.
By comparing the size of the star-forming component of these SFGs to that of their stellar component, we obtain information about the growth of galaxies through cosmic time.

The structure of this paper is as follows. 
In Section~\ref{sec:data and sample}, we present the MIRI and ancillary data used in this study and describe how we built our mass-complete sample of SFGs at $0.1<z<2.5$.
In Section~\ref{sec: MCMC}, we describe the method used to measure the structural parameters of our galaxies.
In Sections~\ref{subsec: sersic} and ~\ref{subsec: sizes}, we present the main results of this study, that is, the rest-MIR \sersic indices and sizes of SFGs and compare them to measurements at rest-optical wavelengths.
In Section~\ref{sec:discussion}, we discuss the implications of our results for galaxy growth models.
Finally, in Section~\ref{sec:summary}, we present our summary.

Throughout this paper, we assume a concordance lambda cold dark matter ($\Lambda$CDM) cosmology, adopting $H_{0}=70$~(km/s)/Mpc, $\Omega_{\rm M}=0.30$ and $\Omega_{\Lambda}=0.70$.
At $z=1$, 1$\arcsec$ corresponds to 8.008\,kpc.
A \citet{Chabrier.2003} initial mass function (IMF) is used for all stellar masses and SFRs quoted in this article.

\section{Data and sample \label{sec:data and sample}}
\subsection{CEERS MIRI}
In this work, we used the MIRI images produced by the CEERS team for their MIRI 1, 2, 5, and 8 fields \citep[][the so-called red MIRI pointings]{Finkelstein.2023}, all in the Extended Groth Strip (EGS) field.
The MIRI 1 and 2 fields benefit from observations with the F770W ($5\sigma\sim25.6\,$AB), F1000W ($\sim24.8$), F1250W ($\sim24.1$), F1500W ($\sim23.6$), F1800W ($\sim22.8$), and F2100W ($\sim22.2$) filters; while the MIRI 5 and 8 fields were observed with the F1000W ($\sim24.6$), F1250W ($\sim23.6$), F1500W ($\sim23.0$), and F1800W ($\sim22.4$) filters \citep{Yang.2023b}. 
Each of these MIRI fields covered an area of approximately 2\,arcmin$^2$ and the full width at half maximum are 0\farcs269 (F770W), 0\farcs328 (F1000W), 0\farcs420 (F1280W), 0\farcs488 (F1500W), 0\farcs591 (F1800W), and 0\farcs674 (F2100W) according to the \textit{JWST} user documentation\footnote{\url{https://jwst-docs.stsci.edu/jwst-mid-infrared-instrument/miri-performance/miri-point-spread-functions}}.
The CEERS MIRI 3, 6, 7, and 9 fields (the so-called blue MIRI pointings) were not used in our analysis because they were taken only with the F560W and F770W filters and thus provide very limited coverage of the rest-MIR emission of high-redshift SFGs. 

The data processing of the MIRI observations is presented in detail in \citet{Yang.2023b}.
Here, we provide only the most relevant properties of this data analysis. 

The MIRI observations were calibrated using the \textit{JWST} calibration pipeline \citep{Bushouse.2022} version 1.7.2 and mostly the default parameters for stages 1 and 2. 
The background was estimated and then subtracted by taking advantage of the multiple images taken in the same bandpass but at different dither positions.
The astrometric correction and stage 3 of the pipeline were then applied to produce the final mosaic, noise, and weight maps.
Their grids are aligned to the \textit{HST} Cosmic Assembly Near-infrared Deep Extragalactic Legacy Survey (CANDELS) v1.9 WFC3 and ACS images \citep{Koekemoer.2011} and use a pixel scale of $0\farcs09$.

Source extraction was performed using the python package \texttt{photutils} in dual mode. 
The detection maps were obtained by creating so-called chi-squared detection images \citep{Szalay.1999}, which are weighted combinations of all MIRI images available over each field, that is, F770W\footnote{The F770W and F2100W images are not available for the MIRI 5 and 8 fields.}, F1000W, F1250W, F1500W, F1800W, and F2100W.
The Kron aperture photometry in each passband was then obtained with \texttt{photutils} in dual mode using these chi-squared-based segmentation maps.
Our final multiwavelength MIRI catalog contains 327 sources with a signal-to-noise ($S/N$) greater than 5 in at least one MIRI bandpass. 
Among these 327 sources, 47, 35, 73, 18, and 65 are detected in 2, 3, 4, 5, and 6 MIRI bandpasses ($73\%$ have $\ge2$ MIRI detections).

\subsection{Ancillary data \label{subsec:ancillary}}

Because the CEERS red MIRI pointings overlap only partially ($\sim45\%$) with the CEERS NIRCam pointings, the optical-to-near-infrared (NIR) photometry of MIRI-detected galaxies was taken from the publicly available multiwavelength catalog of \citet{Stefanon.2017} for the EGS. 
This catalog was built using the \textit{HST} WFC3 and ACS data from CANDELS, the All-wavelength Extended Groth strip International Survey (AEGIS), and the 3D-\textit{HST} programs. 
It is based on detections in the F160W band and reaches a depth of 26.62 AB in this band. 
Supplemented by observations from the Canada-France-Hawaii and Mayall Telescopes as well as from the \textit{Spitzer Space Telescope}, it provides photometry in 23 broadbands from optical to NIR wavelengths. 
Based on this photometry, \citet{Stefanon.2017} also provide robust photometric redshifts ($\Delta z/(1+z)\sim0.02$) and stellar mass measurements for all sources in the catalog.
Finally, when available, \citet{Stefanon.2017} also provide the spectroscopic redshifts of the sources.

We cross-matched our MIRI galaxy catalog with that of \citet{Stefanon.2017}, using search radii of $1\arcsec$.
Among our 327 galaxies, 321 have an optical/NIR counterpart (16 of which have a spectroscopic redshift). 
The six galaxies without a counterpart are so-called \textit{HST}-dark galaxies \citep{Wang.2019}, which means that they are not detected in the \textit{HST} F160W band (on which the catalog of Stefanon et al. is based) but clearly detected by \textit{Spitzer}/IRAC.
As \textit{HST}-dark galaxies are mainly located at $z>3$ \citep{Gomez-Guijarro.2022} and therefore outside our redshift range of interest, we excluded them in the rest of our analysis.

We complemented the optical/NIR photometry with MIR-to-radio photometry from the so-called EGS super-deblended catalog of Le Bail et al. (in prep; see also Liu et al. \citeyear{Liu.2018}, Jin et al. \citeyear{Jin.2018}). 
This super-deblended technique is a prior-based multiwavelength fitting method that optimizes the number of priors fit in each band. 
This catalog provides photometry in the bandpasses of \textit{Spitzer} \citep[24$\,\mu$m from FIDEL;][]{Magnelli.2011}, \textit{Herschel} (100 and 160$\,\mu$m from PEP; Lutz et al. \citeyear{Lutz.2011}; 250, 350, and 500$\,\mu$m from HerMES; Oliver et al. \citeyear{Oliver.2012}), SCUBA2 (850$\,\mu$m from S2CLS; Geach et al. \citeyear{Geach.2017}; 450 and 850$\,\mu$m from Zavala et al. \citeyear{Zavala.2016}), and AzTEC (1.1mm from Aretxaga et al. \citeyear{Aretxaga.2015}). 
Because this catalog is based on optical-to-NIR priors from \citet{Stefanon.2017}, there exists a one-to-one correspondence between these two catalogs. 
Among our 321 MIRI detections with an optical/NIR counterpart, 94 have a $S/N>3$ detection at 24$\,\mu$m and 32 have at least one $S/N>3$ detection at far-infrared (FIR) or (sub)millimeter (submm) wavelengths. 
In absence of FIR or (sub)mm detections, we used $5\sigma$ upper limits for these bands in our spectral energy distribution (SED) fits (see Sect.~\ref{subsec:cigale}).  

Finally, the rest-optical morphology of our MIRI detections was obtained by cross-matching the position of these galaxies with the structural parameter (\sersic index, size, ellipticity) catalog of \citet{Wel.2014}. 
This catalog, which was used for their analysis of the size--mass relation, is based on the CANDELS F125W and F160W imagings as processed by the 3D-$HST$ team. 
These effective radii along the semi-major axis at a rest-frame wavelength of $5000\,\AA$ were calculated by correcting their observed F125W (for galaxies at $z<1.5$) and F160W (for galaxies at $z>1.5$) effective radii for negative color gradients \citep[see equations 1 and 2 in][]{Wel.2014}.  
We cross-matched these catalogs using a search radius of 1\arcsec\ and found counterparts for all our 321 galaxies. 
Above our mass-complete limits (see Sect.~\ref{subsec:final sample}), the number of galaxies with rest-optical structural parameters flagged as ``bad'' in the catalog of \citet{Wel.2014} is negligible (i.e., three out of 72 galaxies).

\subsection{SED modeling with \texttt{CIGALE} \label{subsec:cigale}}
\begin{table*}
\centering
\caption{\texttt{CIGALE} input parameters}
\label{tab:cigale}
\begin{tabular}{llll} \hline\hline
Module & Parameter & Symbol & Values \\
\hline
Star formation history\\ 
\texttt{sfhdelayed}   &  Main stellar population $e$-folding time & $\tau_{\rm main}$ & 0.5, 1, 5\,Gyr\\
                      &  Main stellar population age & $t_{\rm main}$ & 1, 3, 5, 7\,Gyr\\  
                      &  Mass fraction of the late burst & $f_{\rm burst}$ & 0, 0.1, 0.2, 0.3\\
\hline
Simple stellar population\\ 
\texttt{bc03}         & Initial mass function & $-$ & \citet{Chabrier.2003} \\
                      & Metallicity & $Z$ & 0.02 (Solar)\\
\hline
Dust attenuation \\ 
\texttt{dustatt\_calzleit}  & Color excess of stellar continuum & $E(B-V)$ & 0, 0.1, 0.3, 0.5, 0.7, 0.9, 1.1\\
\hline
Dust emission \\ 
\texttt{dl2014}       & PAH mass fraction & $q_{\rm PAH}$ & 0.47, 1.12, 1.77, 2.50, 3.90 \\
                      & Minimum radiation field & $U_{\rm min}$ & 5, 12, 20, 30, 40 \\
                      & Fraction of photodissociation region emission & $\gamma$ & 0.01, 0.05, 0.1, 0.5 \\
\hline
AGN emission \\ 
\texttt{skirtor2016}  & Edge-on optical depth at $9.7\,\mu$m & $\tau_{9.7}$ & 3, 5, 7, 9, 11 \\
                      & Viewing angle & $\theta_{\rm AGN}$ & 30$^\circ$, 70$^\circ$ \\
                      & AGN contribution relative to the dust luminosity & $f_{\rm AGN}$ & 0, 0.05, 0.15, 0.30, 0.45, 0.99\\ 
\hline
\end{tabular}
\begin{flushleft}
{\sc Note.} --- For parameters not listed here, default values have been adopted. We refer the reader to \citet{Boquien.2019} for more details on the \texttt{CIGALE} input parameters.
\end{flushleft}
\end{table*}

\begin{figure}
   \centering
   \includegraphics[angle=0,width=\linewidth]{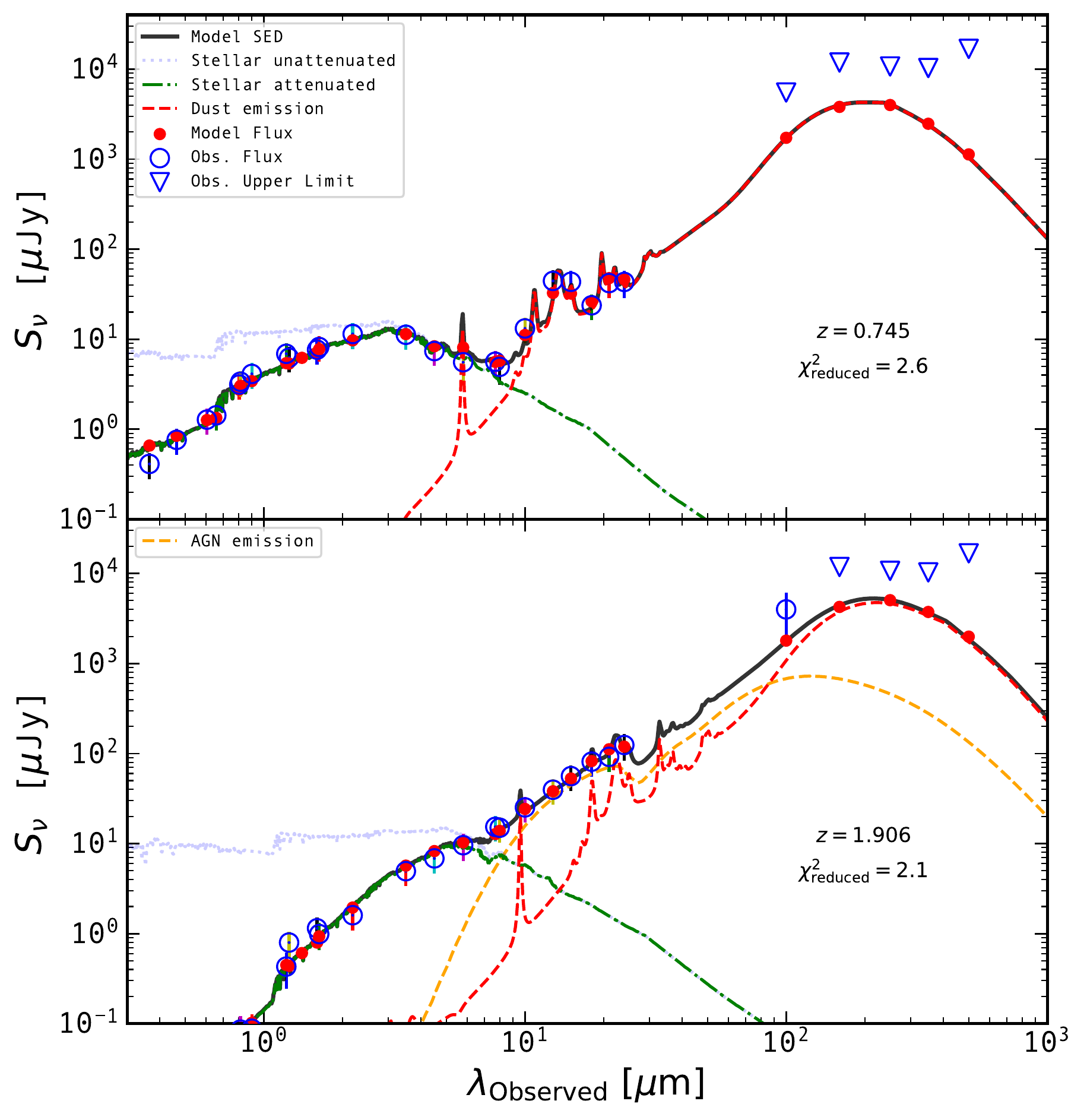}
      \caption{Spectral energy distribution (SED) of two of our galaxies (\textit{upper panel:} ID 12091; \textit{lower panel:} ID 19062) as fit using \texttt{CIGALE}.
      Observed flux densities are shown by opened blue circles, while observed upper limits are shown by downward blue triangles. 
      Black, red, orange, blue, and green lines correspond to the total, dust, AGN, stellar unattenuated and stellar attenuated emission, respectively.
      For the galaxy ID 12091, the AGN contribution (i.e., $L_{\rm AGN}/(L_{\rm AGN}+L_{\rm dust}$) is of $0\%$ and the sharpest MIRI band (shortest wavelength) dominated by dust emission (i.e., $S_{\nu}^{\rm dust}/S_{\nu}^{\rm total} >75\%$) is the 10$\,\mu$m band.
      For the galaxy ID 19062, the AGN contribution is of $30\%$ and none of the MIRI bands is dominated by dust emission.
      }
      \label{fig:cigale}
\end{figure}
\begin{figure*}
   \centering
   \includegraphics[angle=0,width=\linewidth]{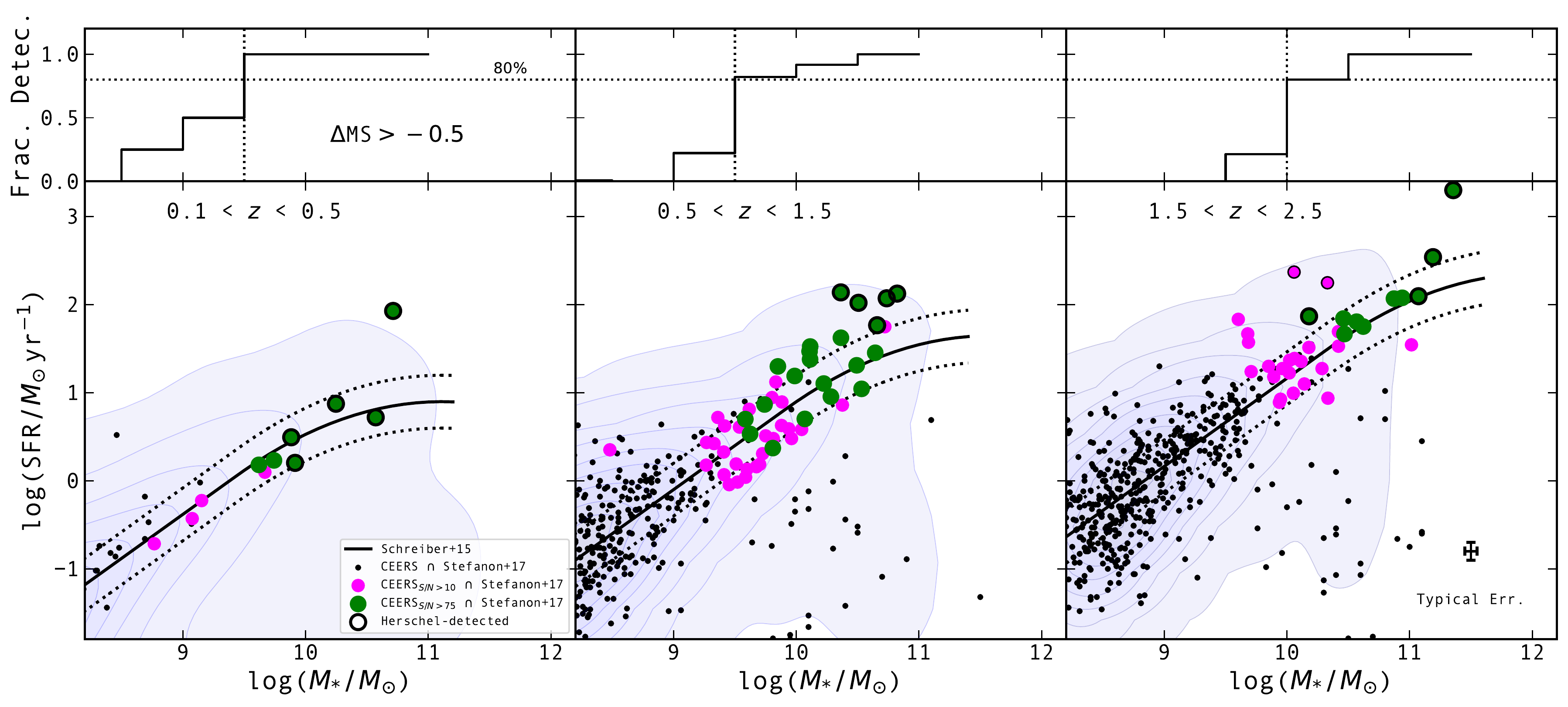}
      \caption{Star formation rate vs. stellar mass distribution in three redshift ranges, i.e., $0.1<z<0.5$, $0.5<z<1.5$, and $1.5<z<2.5$ (\textit{lower panels}). 
      Filled green and magenta circles correspond to rest-MIR detection with $S/N>75$ and $10<S/N<75$, respectively.
      Circles outlined by black edges are detected in the FIR by \textit{Herschel} (i.e., with $S/N_{\rm FIR}>5$, where $S/N_{\rm FIR}$ is the signal-to-noise ratios added in quadrature from 100 to $500\,\mu$m; Le Bail et al. in prep).
      Black dots show galaxies within the four CEERS red MIRI fields but which remained undetected in a MIRI bandpass that is dominated by dust emission (i.e., $S/N<10$ or $S_{\nu}^{\rm dust}/S_{\nu}^{\rm total} <75\%$).
      Shaded regions correspond to the distribution of all sources in the EGS, plotted using a kernel density estimator.
      The solid and dotted black lines show the MS of SFGs and its dispersion as inferred by \citet{Schreiber.2015}.
      (\textit{upper panels}) Fraction of SFGs (i.e., $\Delta MS>-0.5$) as a function of stellar mass which are within the four CEERS red MIRI fields and are in our sample (i.e., that have at least one $S/N>10$ MIRI detection dominated by dust emission). 
      Our sample is $80\%$ mass-complete down to $M_{\ast}\sim10^{9.5}, 10^{9.5},$ and $10^{10}\,M_{\odot}$ at $z\sim0.3$, 1.0, and 2.0, respectively.
      A typical $1\sigma$ error bar for individual objects is shown in the right panel.
      }
      \label{fig:sfr vs mstar}
\end{figure*}
\begin{figure}
   \centering
   \includegraphics[angle=0,width=\linewidth]{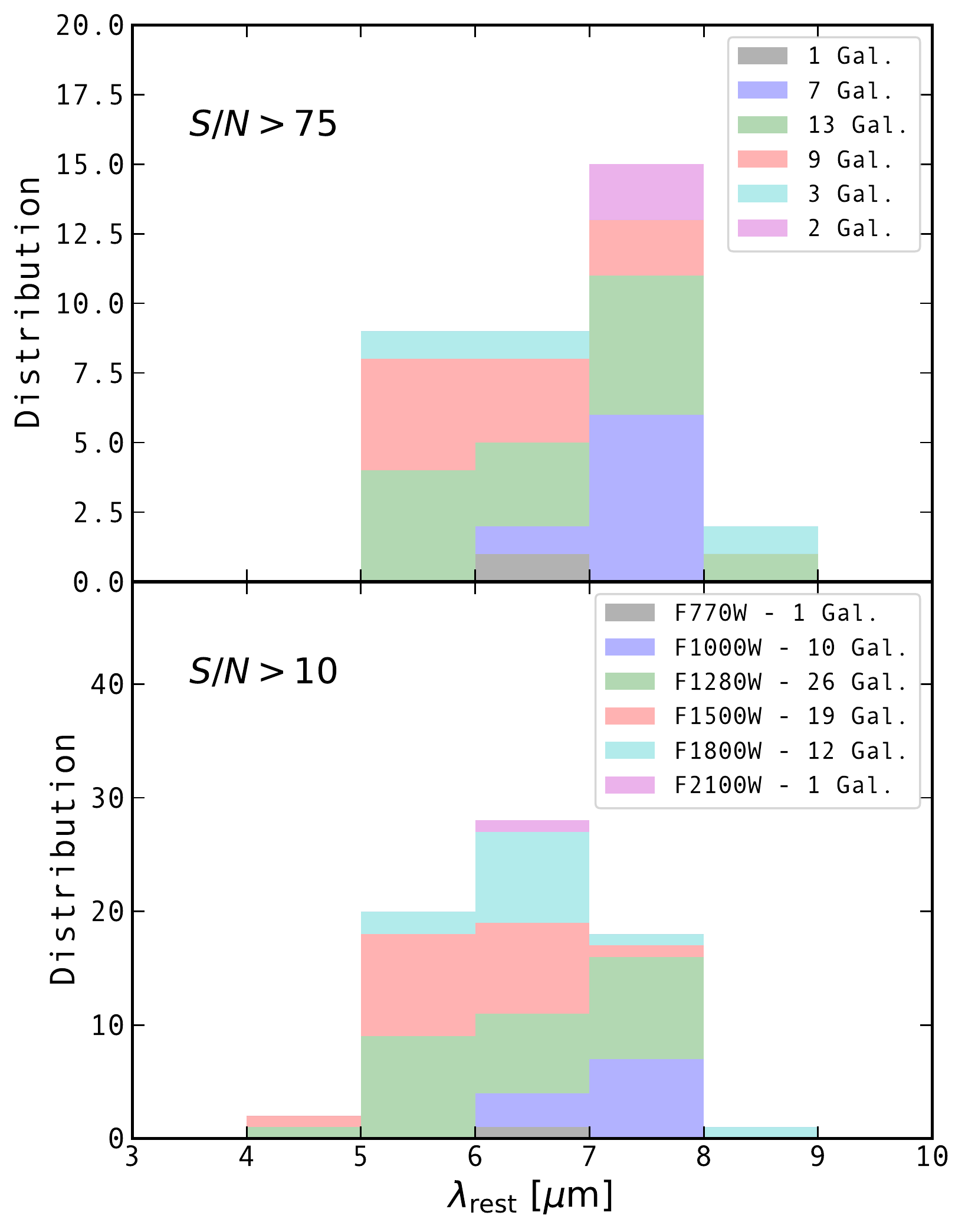}
      \caption{Distributions of rest-MIR wavelengths probed by the sharpest MIRI band (shortest wavelength) with a detection dominated by dust emission, for our $S/N>75$ (\textit{upper panel}) and $S/N>10$ (\textit{lower panel}) samples.
      The colors of these stacked distributions correspond to the MIRI band used.
      The total number of galaxies in a given MIRI band is shown in the legend of each panel.
      }
      \label{fig:MIRI band}
\end{figure}
To construct our final sample (see Sect.~\ref{subsec:final sample}) from these 321 MIRI detections with counterparts in the catalog of \citet{Stefanon.2017}, we first fit their photometry from optical to (sub)mm wavelengths using \texttt{CIGALE} \citep{Boquien.2019,Yang.2019}. 
In doing so, we fixed their redshifts to those inferred in Stefanon et al. and used a delayed star formation history with an optional exponential burst, the \citet{Bruzual.2003} stellar model, the \citet{Calzetti.2001} attenuation law, the \citet{Draine.2014} dust emission model (imposing energy balance) and the \citet{Stalevski.2016} Active Galatic Nuclei (AGN type 1 and 2) SED model.
As we included the AGN and dust emission modules of \texttt{CIGALE} in our fits, this requires the generation of a very large number of models per galaxy.  
The parameter grids adopted here (Tab.~\ref{tab:cigale}) have therefore been selected to obtain accurate fits, while keeping the number of models generated to a reasonable level.
During the fit a minimum uncertainty of $10\%$ is used to account for flux calibration uncertainties \citep{Boquien.2019}.
We inspected all 321 fits and found none significantly bad, consistent with the absence of outliers in their reduced $\chi^2$ distribution. 
Figure~\ref{fig:cigale} displays two examples of these fits, one for which the AGN contribution --i.e., $L_{\rm AGN}/(L_{\rm AGN}+L_{\rm dust})$, where $L_{\rm AGN}$ and $L_{\rm dust}$ are the dust luminosity of the AGN and the dust luminosity of the host galaxy, respectively-- is of $0\%$ and one for which the AGN contribution is of $30\%$.

Based on these \texttt{CIGALE} best fits, we first excluded from our sample 60 galaxies with a significant contribution from an AGN (i.e., $L_{\rm AGN}/(L_{\rm AGN}+L_{\rm dust}) \equiv f_{\rm AGN}> 10\%$).
The presence of these (obscured) AGNs was assessed by \texttt{CIGALE} thanks to their characteristic power-law rest-MIR emission \citep[see lower panel of Fig.~\ref{fig:cigale}; see also, e.g.,][]{Donley.2012,Yang.2023}.
The exclusion of these (obscured) AGNs prevents contamination in our morphological analysis, as their emission dominates in the rest-MIR, with $L_{\rm AGN}[5-25\mu{\rm m}]/(L_{\rm AGN}[5-25\mu{\rm m}]+L_{\rm dust}[5-25\mu{\rm m}] \gtrsim30\%$ (see the lower panel of Fig.~\ref{fig:cigale}). 
The fraction of (obscured) AGNs found in our study is consistent with the results of \citet{Yang.2023}.

Using our \texttt{CIGALE} best fits, we then excluded from our sample 57 galaxies that are classified as quiescent based on their rest-optical colors and using a standard $UVJ$ selection method \citep[e.g.,][]{Mortlock.2014}.
We note that one source was formally classified as quiescent using this $UVJ$ selection but nevertheless detected in the MIR-to-FIR catalog of Le Bail et al. (in prep).
This misclassified galaxy has been reintegrated in our sample of SFGs. 

Finally, using our \texttt{CIGALE} best fits, we defined for each remaining galaxy which MIRI bands (if any) were dominated by dust emission, that is, $S_{\nu}^{\rm dust}/S_{\nu}^{\rm total} >75\%$. 
For a given galaxy, when more than one MIRI band  was dominated by dust emission, we adopted the band with the shortest wavelength (i.e., with the best angular resolution) for our rest-MIR structural measurements\footnote{Adopting instead the band with the longest wavelength would not change the results presented in this paper.}. 
In the following, we refer to this band as the sharpest MIRI band (i.e., shortest wavelength) dominated by dust emission.
For example, for the galaxy ID 12091 shown in the upper panel of Fig.~\ref{fig:cigale}, this corresponds to the MIRI 10$\,\mu$m band (i.e., F1000W). 
Our sample contains 95 SFGs at $0.1<z<2.5$ that have a MIRI band dominated by dust emission and a detection significance in that band of $S/N>10$ (i.e., the minimum detection significance required to accurately measure the rest-MIR size of a galaxy fixing its \sersic index to 1; see Sect.~\ref{sec: MCMC}); seven out of these 95 galaxies have a spectroscopic redshift. 
This number drops to 38 if we only consider bands with $S/N>75$ (i.e., the minimum detection significance required to accurately measure both their rest-MIR sizes and \sersic indices; see Sect.~\ref{sec: MCMC}); two out of these 38 galaxies have a spectroscopic redshift.

In the rest of this analysis, the SFRs, stellar masses, and associated uncertainties of these 95 SFGs are taken from the \texttt{CIGALE} fits.
We checked that these SFRs were consistent ($\sigma\sim0.2\,$dex) with those inferred by scaling the MS template of \citet{Elbaz.2011} to the MIRI or \textit{Herschel} flux densities of these galaxies.
There are, however, three galaxies detected by \textit{Herschel} for which \texttt{CIGALE} underestimated their SFR by a factor of three compared to those deduced from the FIR.
For these galaxies, we used their \textit{Herschel}-based SFRs instead.

\subsection{Final sample \label{subsec:final sample}}
The SFR and stellar mass distributions of our sample of 95 SFGs that have a MIRI band dominated by dust emission and a detection significance in that band of $S/N>10$ are shown in Fig.~\ref{fig:sfr vs mstar}. 
Our galaxies clearly follow the MS of SFGs but also include some starbursts located above this sequence. 
In the upper panels of Fig.~\ref{fig:sfr vs mstar}, we studied the fraction of SFGs (i.e., $\Delta MS\equiv {\rm log}(SFR_{\rm gal.} (M_\ast,z)/SFR_{\rm MS}(M_\ast,z))>-0.5$) which are within the four CEERS red MIRI fields and are in our sample, that is, with at least one $S/N>10$ MIRI detection dominated by dust emission. 
From this analysis, we find that even with our relatively strict rest-MIR detectability of $S/N>10$, our sample is $80\%$ mass-complete down to $M_{\ast}\sim10^{9.5}, 10^{9.5},$ and $10^{10}\,M_{\odot}$ at $z\sim0.3$, 1.0, and 2.0, respectively. 
In fact, the ten SFGs (i.e., $\Delta MS>-0.5$) above these stellar masses that do not appear in our sample are AGNs that we excluded because a detailed analysis of their host galaxy's rest-MIR morphology was practically impossible, $L_{\rm AGN}[5-25\mu{\rm m}]/(L_{\rm AGN}[5-25\mu{\rm m}]+L_{\rm dust}[5-25\mu{\rm m}] \gtrsim30\%$.
Our sample should therefore be representative of the population of SFGs at intermediate mass ($\sim10^{10-11}\,M_\odot$).
At higher stellar masses, although our sample is complete, it still suffers from cosmic variance and misses the rare massive SFGs ($\gtrsim10^{11}\,M_\odot$) due to the small areal coverage of our MIRI observations.

For ease of interpretation, in the remainder of the analysis we restricted our sample to galaxies above these mass-complete limits of $M_{\ast}\sim10^{9.5}, 10^{9.5},$ and $10^{10}\,M_{\odot}$ at $z\sim0.3$, 1.0, and 2.0, respectively. 
This sample, which originally contains 72 galaxies with robust rest-MIR structural measurements, drops to 69 galaxies when we consider only galaxies with rest-optical structural measurements not flagged as bad in the catalog of \citet{Wel.2014}.
This mass-complete sample of 69 SFGs with robust rest-optical and rest-MIR structural measurements is our final sample.
This final sample is, when needed, further segregated according to the MIRI detection significance (69 with $S/N>10$ and 35 with $S/N>75$).

Figure~\ref{fig:MIRI band} shows the distribution of rest-MIR wavelengths used for the morphological analysis of this final sample, that is, the wavelengths probed by their sharpest MIRI band dominated by dust emission.
These rest-MIR wavelengths correspond mainly to the PAH 6.2 and 7.7$\,\mu$m features, which implies that for galaxies located at a higher redshift, we use MIRI bands at longer (redder) effective wavelengths.
This does not seem to affect our results, as none of the trends observed in the paper depend on the redshift of our galaxies.

Finally, using the \texttt{CIGALE} best fits of these 69 SFGs, we studied the fraction of their SFR that is obscured by dust, $\rm{SFR}_{\rm IR}/({\rm SFR}_{\rm UV}+{\rm SFR}_{\rm IR})$, where $\rm{SFR}_{\rm IR}$ is the dust-obscured SFR and is given by ${\rm SFR}_{\rm IR} [M_{\odot}\,{\rm yr^{-1}}]=1.09\times10^{-10}\times L_{\rm dust}[8-1000\,\mu{\rm m}; L_{\odot}]$, while $\rm{SFR}_{\rm UV}$ is the unobscured SFR and is obtained from the rest-frame luminosity at 2300\,\AA, $L_{\rm UV}$, as ${\rm SFR}_{\rm UV} [M_{\odot}\,{\rm yr^{-1}}]=3.6\times10^{-10}\times L_{\rm UV}[L_{\odot}]$ \citep{Kennicutt.1998}.
We found a median value of $0.89_{-0.20}^{+0.06}$ for $\rm{SFR}_{\rm IR}/({\rm SFR}_{\rm UV}+{\rm SFR}_{\rm IR})$ (the range correspond to the 16th and 84th percentiles). 
This is consistent with literature results for the same stellar mass range as that probed by our final sample \citep[see, e.g.,][]{Whitaker.2017,Shen.2023}.
In what follows, we therefore consider the star-forming component of our galaxies to be accurately traced by their dust-obscured star-forming component.
\section{Structural parameter measurements \label{sec: MCMC}}
To constrain the rest-MIR structural parameters of our galaxies, we performed a standard Bayesian analysis using the python Markov chain Monte Carlo (MCMC) package \texttt{emcee} \citep{Foreman-Mackey.2013}.
For each galaxy, we started from the analytical formula of a \sersic profile (defined by its \sersic index, effective radius, ellipticity, and position), convolved the model to be tested with the point spread function (PSF) of the corresponding MIRI band, and finally calculated the likelihood of this model (i.e., $\exp{(-\chi^2/2)}$) by comparison with the real image assuming that the noise in the MIRI image is Gaussian.
These noise maps are based on the weight maps generated by the \textit{JWST} pipeline but we multiplied them by a factor $\sim2-3$ (depending on the field and band) to account for pixel-correlated noise \citep{Yang.2023b}.
Given that the number density of stars in our MIRI images (especially in the longest bandpass) is relatively low and that we only have four MIRI fields covering a total of $\sim\,$8\,arcmin$^2$, creating an empirical PSF is impractical. 
As an alternative, we used PSFs generated using \texttt{WebbPSF} \citep{Perrin.2014}, assuming a pixel scale of 0\farcs09 and accounting for the position angle of the different CEERS exposures. 
   \begin{figure*}
   \centering
   \includegraphics[angle=0,width=\linewidth]{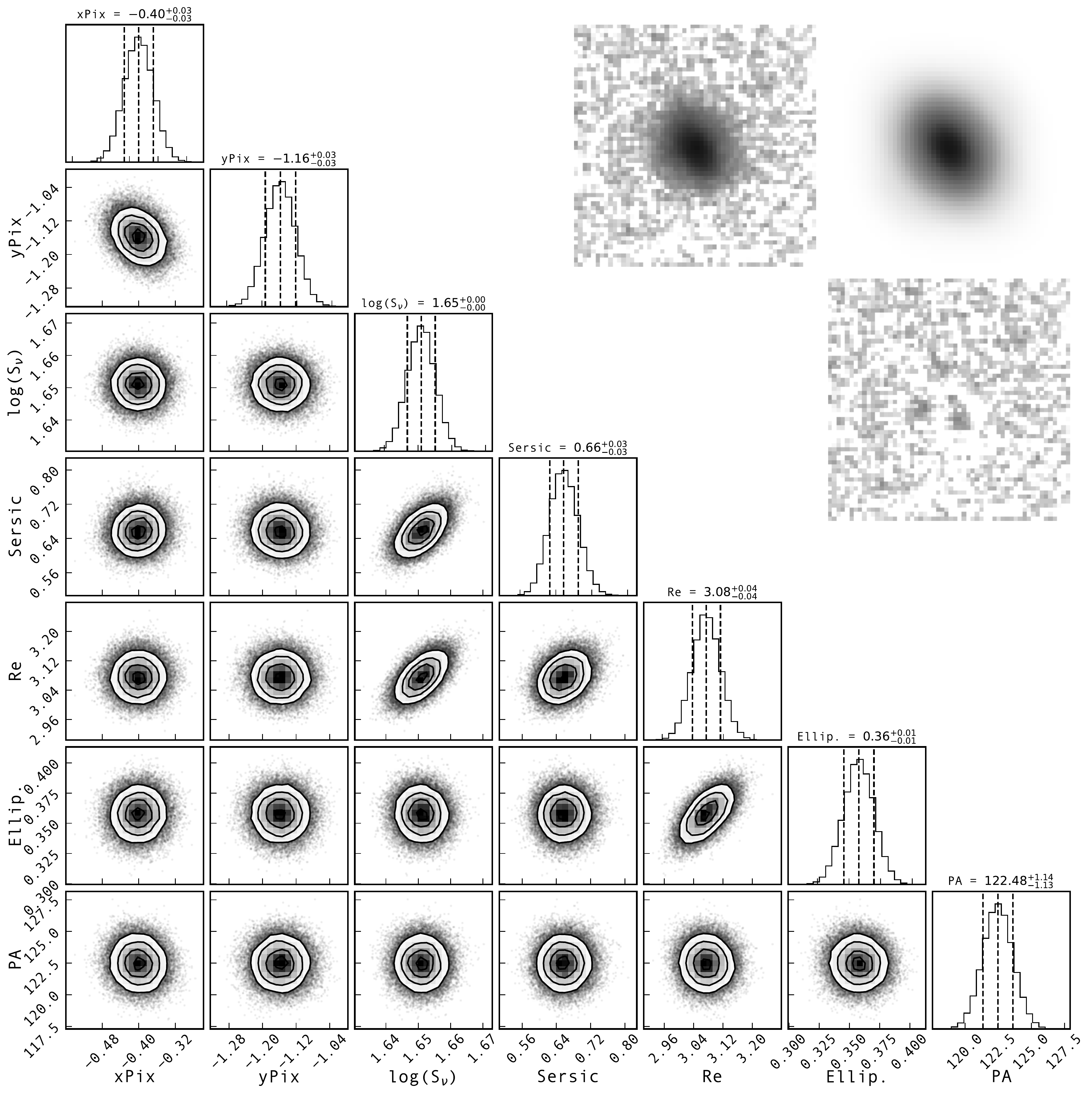}
      \caption{Probability distributions of the structural parameters of one of our galaxies (ID 12091).
      From left to right, $x$ and $y$-offset from the center of the $5\arcsec\times5\arcsec$ cutout ([pixel] with a pixel scale of 0\farcs09), total flux ([log$(\mu Jy)$]), \sersic index, effective radius ([kpc]), ellipticity, and position angle ([degree]).
      This galaxy is bright enough (i.e., $S/N> 75$) that we let its \sersic index as a free parameter.
      The dashed vertical lines show the 16th, 50th, and 84th percentiles of each distribution.
      The upper right panels show the original, model, and residual (original minus model) images of this galaxy in the 10$\,\mu$m band.
      }
      \label{fig:mcmc}
   \end{figure*}
   \begin{figure*}
   \centering
   \includegraphics[angle=0,width=0.85\linewidth]{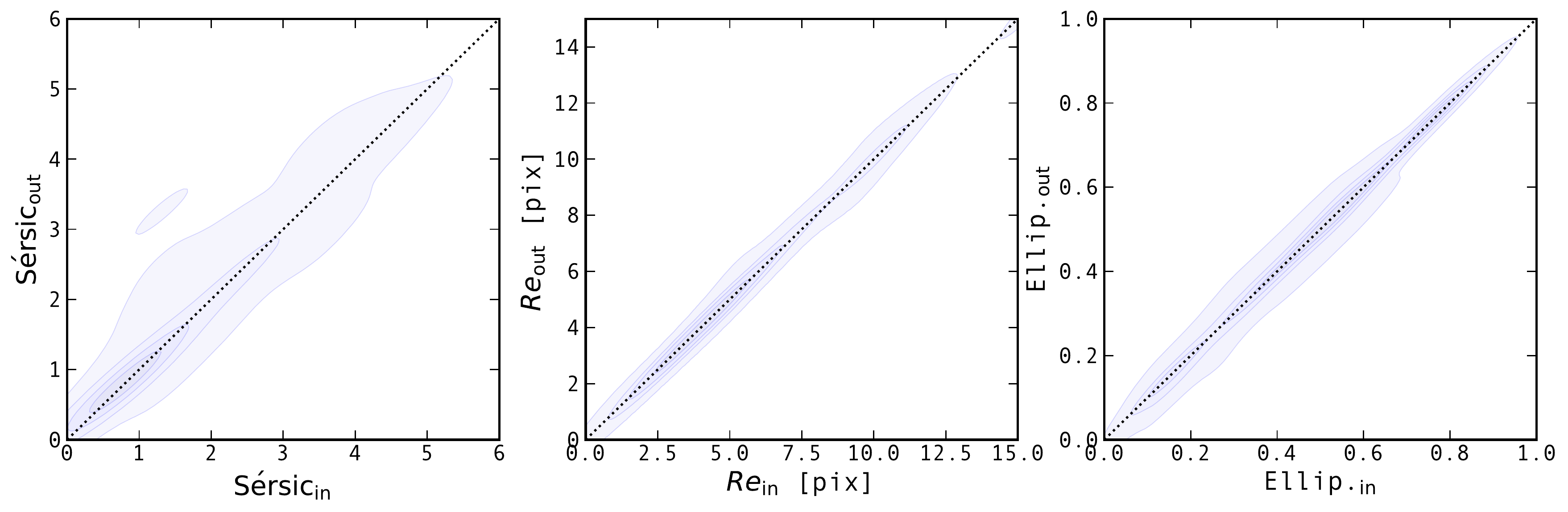}
      \includegraphics[angle=0,width=0.85\linewidth]{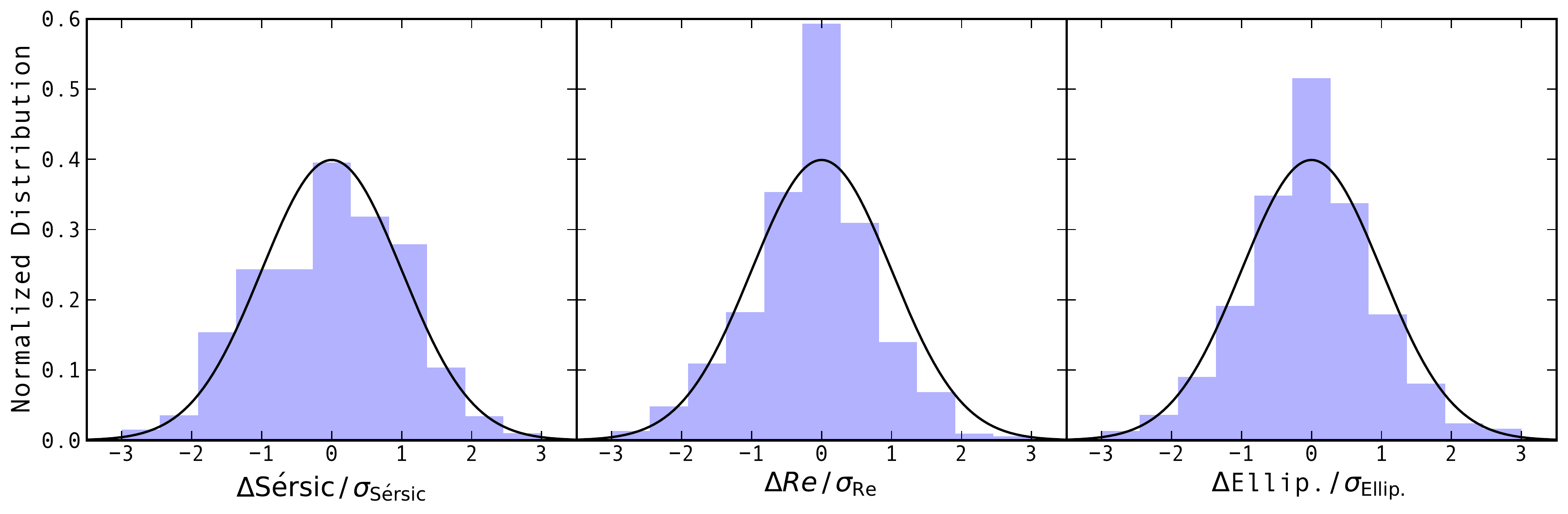}
      \includegraphics[angle=0,width=0.566\linewidth]{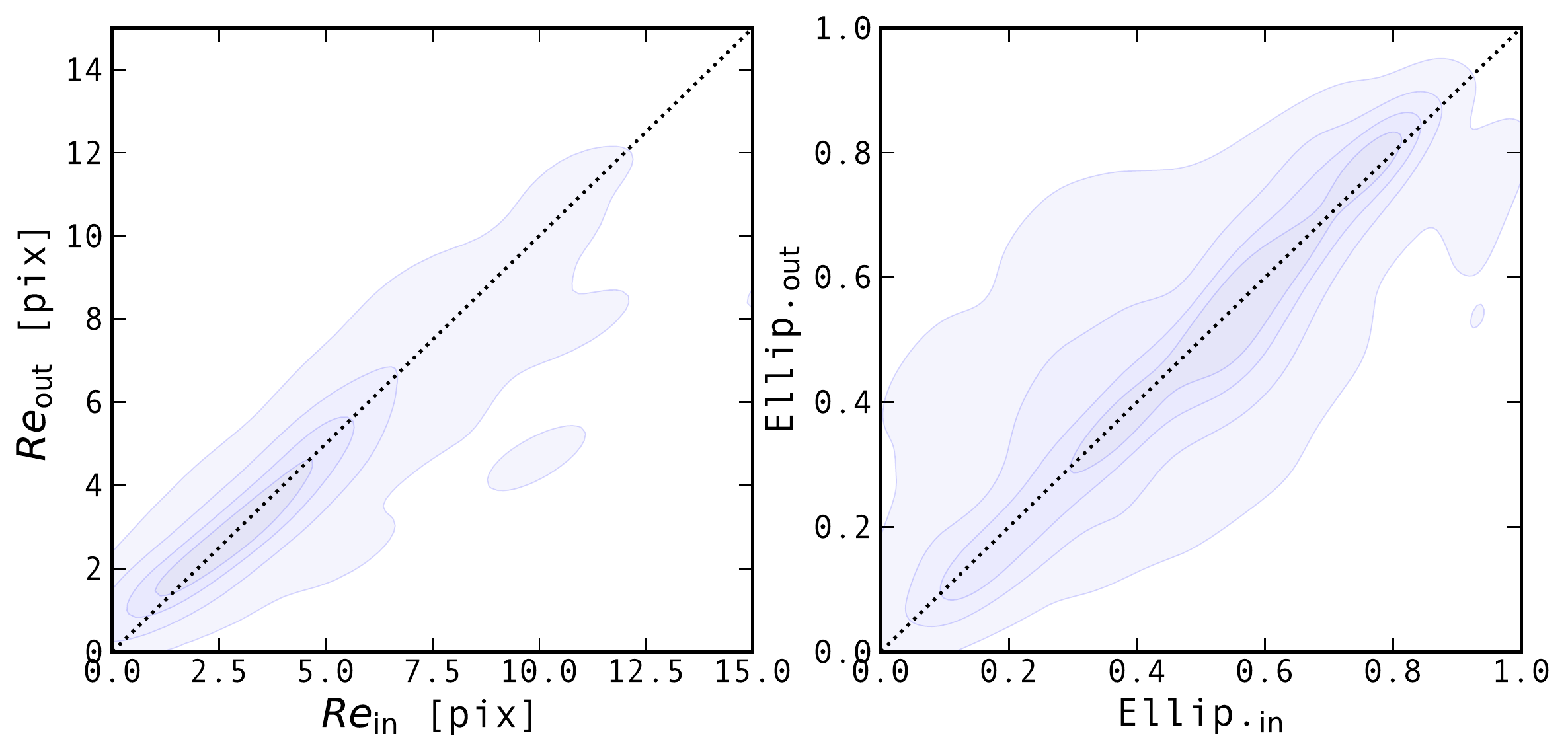}
      \includegraphics[angle=0,width=0.566\linewidth]{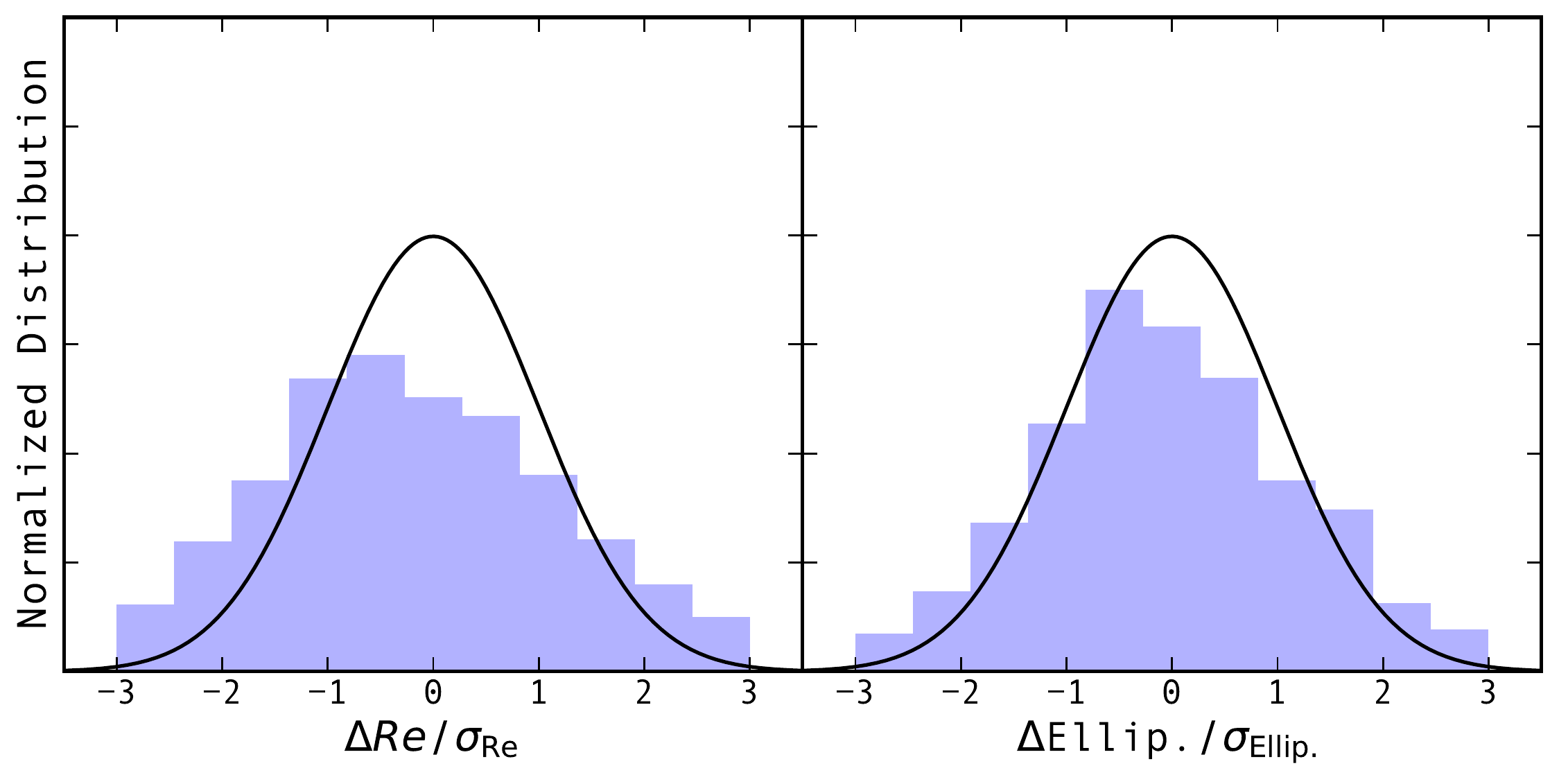}
      \caption{Quality of our structural parameter measurements as inferred using Monte-Carlo simulations.
      (\textit{1st row}) Comparison between the intrinsic (``in'') and measured (``out'') \sersic indices, effective radii, and ellipticities. 
      Shaded regions correspond to the distribution of all 3,500 simulated data points, plotted using a kernel density estimator. 
      Dotted lines show the one-to-one relation.
      (\textit{2nd row}) Distribution of the true (i.e., $\Delta \rm{ Param} = \rm{Param_{in}} - \rm{Param_{out}}$) to estimated (i.e., $\sigma_{\rm Param}$) error ratio (blue histogram).
      The solid line shows a Gaussian distribution centered on zero and with a dispersion of one, i.e., the distribution that should be followed by the blue histogram if the estimated errors were statistically accurate.
      In the first two rows, mock galaxies were simulated using the observed rest-MIR flux density of each galaxy with $S/N>75$ in our sample and their respective rest-optical \sersic indices and sizes.
      Structural parameters were then retrieved using the MCMC approach described in Sect.~\ref{sec: MCMC}.
      (\textit{3rd row} and \textit{4th row}) Same as the first two rows but for simulations performed using the observed rest-MIR flux density of galaxies with $10<S/N<75$ and their respective rest-optical \sersic indices and sizes. 
      Structural parameters were then retrieved using the MCMC approach described in Sect.~\ref{sec: MCMC} but fixing the \sersic index to 1.
      }
      \label{fig:simu}
   \end{figure*}
To perform our fits, we used cutouts of $5\arcsec\times5\arcsec$ centered on each of our galaxies. 
Our complete \sersic model has seven parameters: the $x$ and $y$ offset from the center of the cutout, the total flux ($S_\nu$), the \sersic index ($n$), the effective radius ($Re$), the ellipticity (i.e., $\varepsilon = 1-b/a$), and the position angle (PA).
We adopted the effective radius along the semi-major axis to avoid inclination and projection effects.
To account for the impact of neighboring sources, we performed our analysis three times. 
In the first pass, we fit all sources with $S/N > 10$ without considering neighboring sources and fixing the \sersic index of sources with $10 < S/N < 75$ (see below) to 1. 
In the second and third passes, we fit each cutout after subtracting neighboring sources using their structural parameters found in the previous pass. 
The models converge after three passes. 
   \begin{figure*}
   \centering
   \includegraphics[angle=0,width=0.495\linewidth]{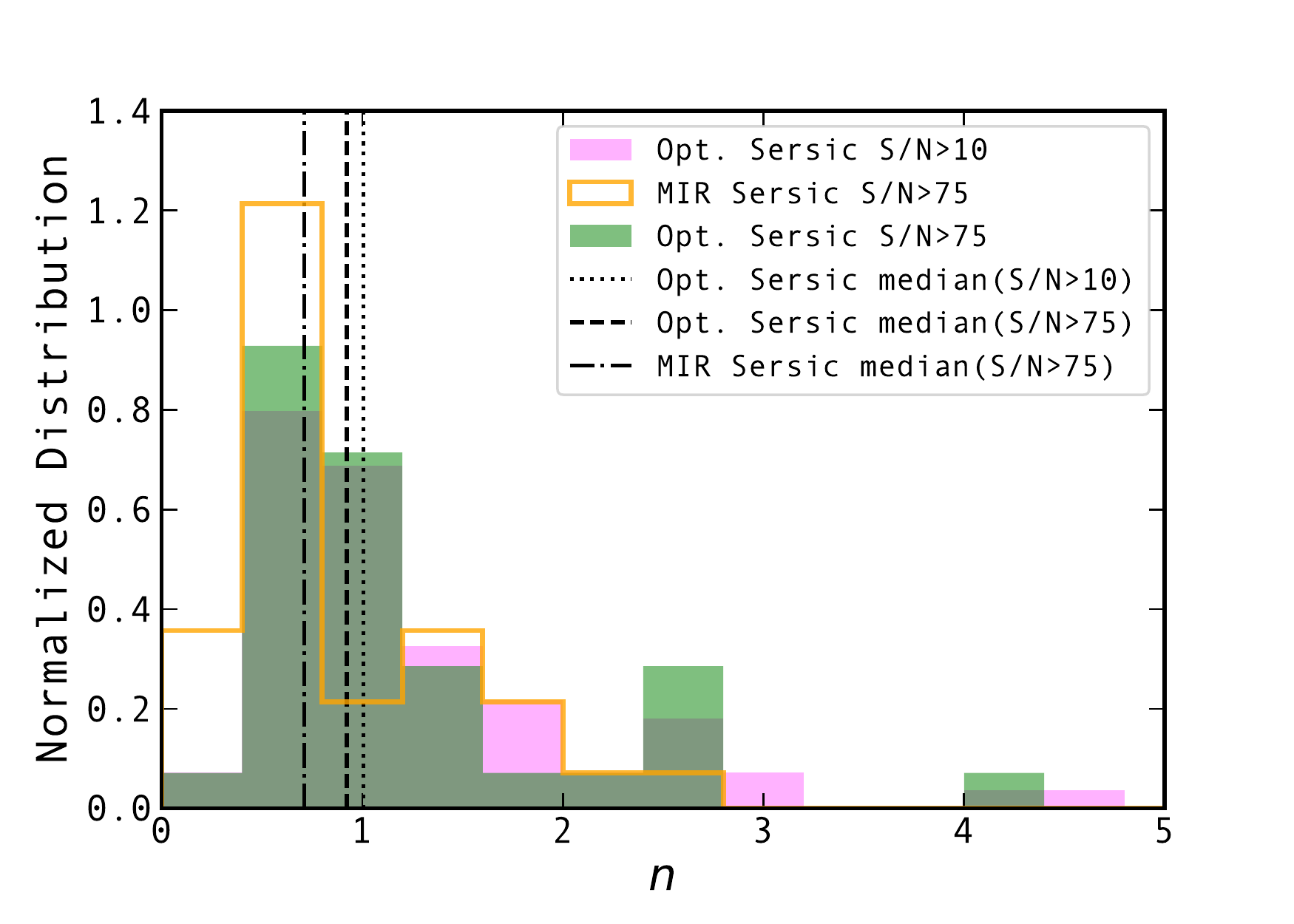}
          \includegraphics[angle=0,width=0.495\linewidth]{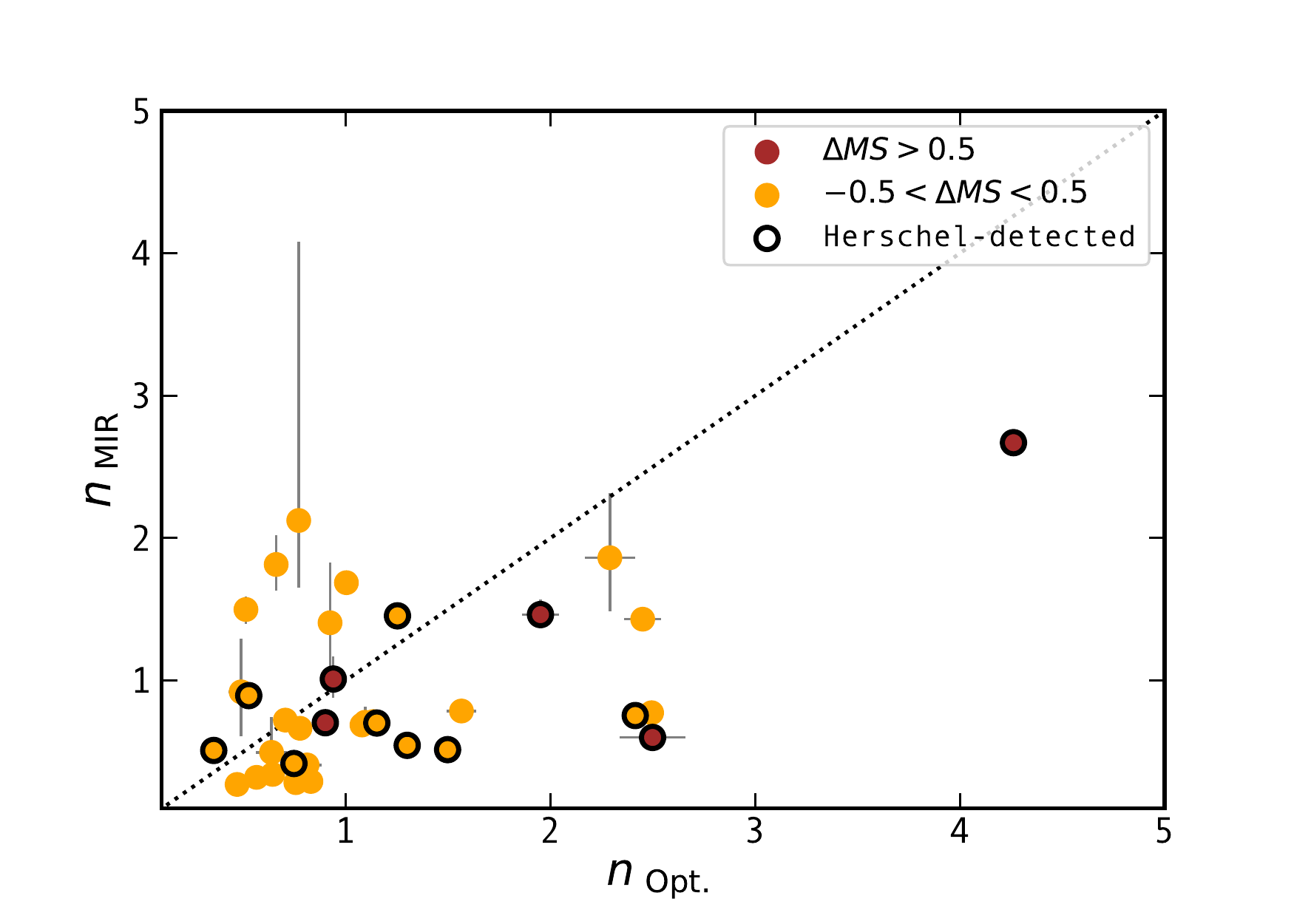}

      \caption{\sersic indices of our final sample.
      (\textit{left}) Distributions of the rest-optical and rest-MIR \sersic indices for our final sample. 
      The green and magenta distributions correspond to the rest-optical \sersic indices of our $S/N>75$ and $S/N>10$ samples, respectively.
      The orange distribution corresponds to the rest-MIR \sersic indices of our $S/N>75$ sample.
      The vertical dashed, dotted and dash-dotted lines show the median \sersic indices of these distributions.
      (\textit{right}) Comparison of the rest-MIR and rest-optical \sersic indices of the 35 SFGs with $S/N>75$ in our final sample.
      Circles are color-coded by the distance of each galaxy to the MS, i.e., $-0.5 < \Delta MS < 0.5$ (orange), and $\Delta MS > 0.5$ (brown).
      Circles outlined by black edges are detected in the FIR by \textit{Herschel}.
      }
      \label{fig:sersic distri}
   \end{figure*}
   
An example of this structural analysis for one of our bright ($S/N>75$) galaxies is shown in Fig.~\ref{fig:mcmc}. 
It illustrates the advantage of this MCMC approach from which we can recover the complete posterior probability distribution and analyze covariance between parameters. 
In what follows, we use the 50th percentile as the best fit and the 16th and 84th percentiles as errors. 
We visually inspected all models (i.e., cutouts and residual cutouts) and found no problematic fit.
   
We validated our structural parameter measurements using Monte Carlo simulations. 
These simulations are based on the observed flux densities of our galaxies in the sharpest MIRI bandpass dominated by dust emission and their rest-optical structural parameters as taken from the structural parameter catalogs of \citet[][; see Sect.~\ref{subsec:ancillary}]{Wel.2014}. 
For each galaxy, we proceeded as follows: from its MIRI flux density and rest-optical structural parameters, we created its mock light profile and introduced it in 100 different positions of the MIRI image, avoiding real sources using the corresponding segmentation map; then we retrieved the structural parameters of these 100 light profiles and compared the intrinsic and measured structural parameters. 

With this set of simulations, we first assessed the lowest $S/N$ below which accurate \sersic index measurements could be obtained. 
We find that at $S/N<75$ the quality of our measurements significantly degraded, driven especially by inaccurate \sersic index measurements (relative error $>50\%$). 
In what follows, we thus restricted our full structural analysis (i.e., $n$, $Re$, and $\varepsilon$) to our brightest 35 sources with $S/N> 75$. 
We note that such restriction is fully consistent with the analysis of \citet{Wel.2014}, who also restricted their \sersic index measurements to sources 2 to 3 magnitude brighter than their 5$\sigma$ detection threshold (i.e., $S/N>30-75$). 

In the first two rows of Fig.~\ref{fig:simu}, we present the results of this first set of simulations, restricted to galaxies with $S/N>75$. 
Our method accurately retrieves all three structural parameters across the entire dynamic range tested by our simulations, with a median in-to-out parameter ratio of $1.00_{-0.16}^{+0.07}$, $1.00_{-0.03}^{+0.03}$, and $1.00_{-0.05}^{+0.07}$ (the ranges correspond to the 16th and 84th percentiles) for $n$, $Re$, and $\varepsilon$, respectively. 
Furthermore, not only our method accurately retrieves these structural parameters but it also provides accurate error measurements as specifically demonstrated by the second row of Fig.~\ref{fig:simu}. 
Indeed, the true (i.e., $\Delta \rm{ Param} = \rm{Param_{in}} - \rm{Param_{out}}$) to estimated (i.e., $\sigma_{\rm Param}$) error ratio follows a Gaussian distribution centered at zero and with a dispersion of one, as expected in the case of accurate error measurements.

To extend our morphological analysis to galaxies with $S/N<75$, we limited our structural parameter measurements to their effective radii and ellipticities by fixing the value of their \sersic index.
We decided to fix $n_{\rm MIR}$ to a value of 1 (exponential light profile\footnote{Fixing instead the \sersic index of each of our galaxies to that observed in the rest-optical would not change any of the results presented in this paper.}) because SFGs are known to be dominated by exponential light profiles, as demonstrated by the rest-optical and rest-MIR \sersic index distributions of our sample (Fig.~\ref{fig:sersic distri}; Sect~\ref{subsec: sersic}).
   
Using the same set of simulations, we then evaluated the quality of our radius and ellipticity measurements when fixing $n_{\rm MIR}$ to 1. 
First, we found that at $S/N<10$ the quality of our $Re$ measurements significantly degraded and so restricted our partial structural parameter analysis (i.e., $Re$ and $\varepsilon$ with $n_{\rm MIR}=1$) to sources with $S/N>10$.
The quality of our partial structural parameter analysis for galaxies with $S/N> 10$ is shown in the bottom two rows of Fig.~\ref{fig:simu}. 
Our radius and ellipticity measurements are accurate with a median in-to-out parameter ratio of $0.99_{-0.18}^{+0.22}$ and $0.99_{-0.27}^{+0.28}$, respectively. 
However, these measurements are characterized by larger dispersions due in part to the lower $S/N$ but also to our $n_{\rm MIR}=1$ assumption, which necessarily propagates into measurement inaccuracies ($n_{\rm in} \ne 1$). 
Our errors are also affected by this assumption, with the distributions of the true to estimated errors ratio following Gaussian distributions centered on 0 but with dispersions of $\sim1.1$. 
We conclude that our errors are underestimated by $\sim10\%$ (see also Sect.~\ref{subsec: sersic}) because they do not account for the systematic error introduced by our $n_{\rm MIR}=1$ assumption. 
However, this effect remains relatively moderate because the galaxies in our sample have a median rest-optical \sersic index close to a value of one. 
In what follows, we decided to simply increase by $10\%$ the errors associated with our partial structural parameter analysis.

\section{Results}
\subsection{Rest-MIR \sersic indices \label{subsec: sersic}}
   \begin{figure}
   \centering
   \includegraphics[angle=0,width=\linewidth]{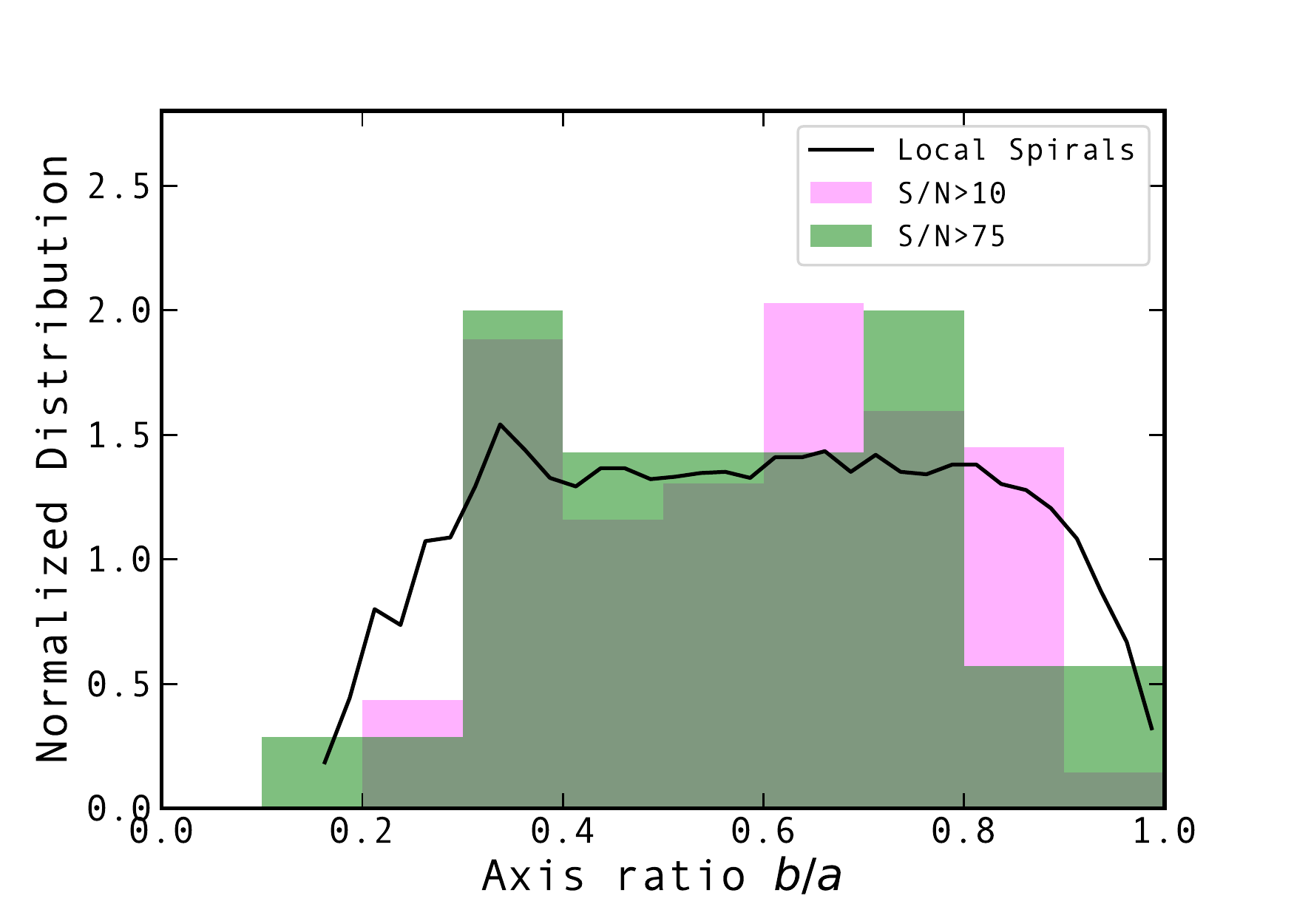}
      \caption{Rest-MIR axis ratio (i.e., $b/a = 1- \varepsilon$) distributions of our final sample.
      The green and magenta distributions correspond to our $S/N>75$ and $S/N>10$ samples, respectively.
      The black solid line shows the rest-optical axis ratio distribution of local spirals \citep[][]{Rodriguez.2013}.
      }
      \label{fig: Ellipticity}
   \end{figure}
      \begin{figure*}
   \centering
   \includegraphics[angle=0,width=\linewidth]{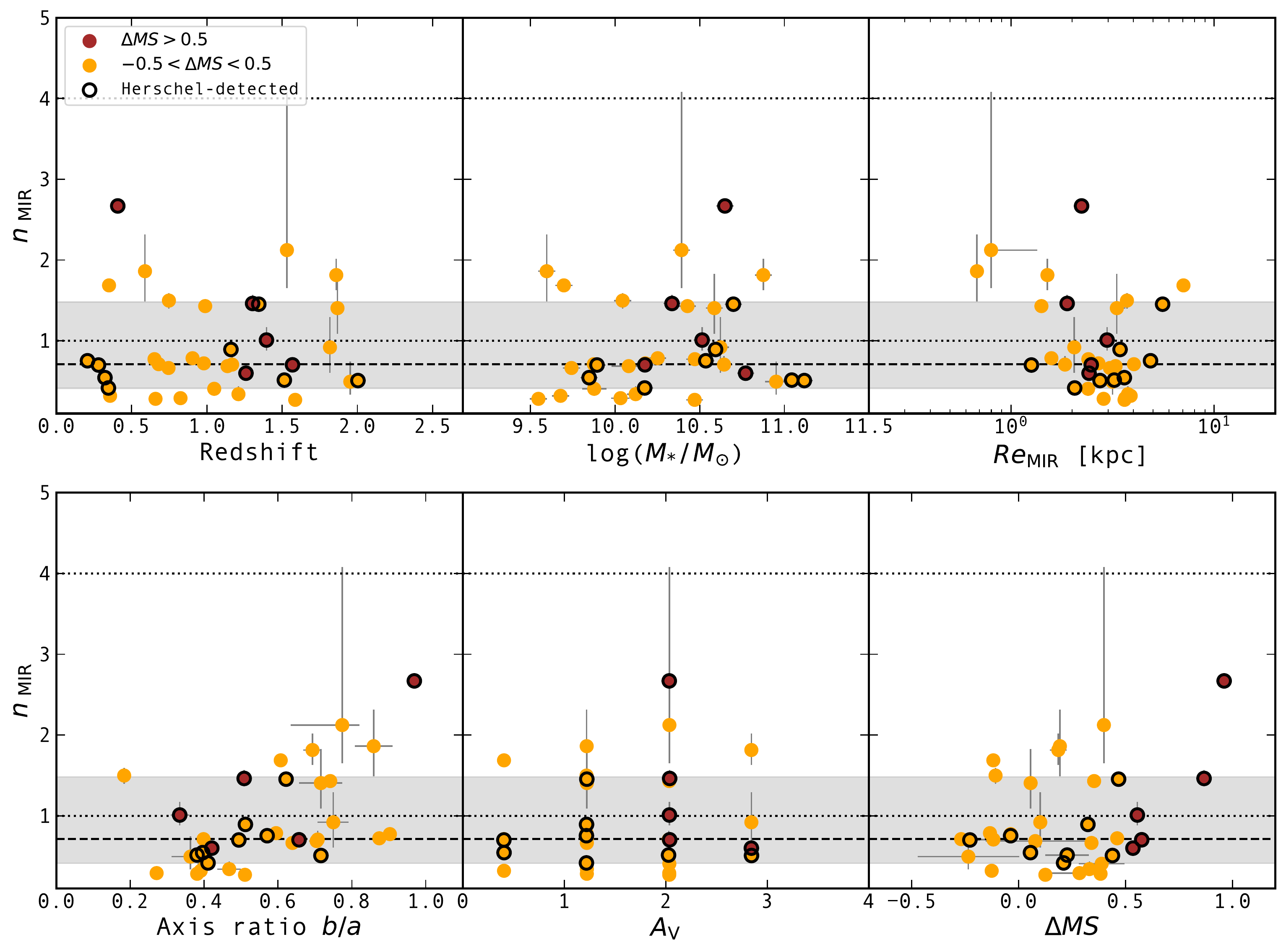}
      \caption{Rest-MIR \sersic indices of the 35 SFGs with $S/N>75$ in our final sample as a function of their redshifts, stellar masses, rest-MIR effective radii, axis ratio, dust attenuation, and distances to the MS.
      Circles are color-coded by the distance of each galaxy to the MS, i.e., $-0.5 < \Delta MS < 0.5$ (orange), and $\Delta MS > 0.5$ (brown).
      Circles outlined by black edges are detected in the FIR by \textit{Herschel}.
      The shaded regions show the 16th to 84th percentiles of the rest-MIR \sersic index distribution, while the horizontal dashed lines represent its 50th percentile.
      The horizontal dotted lines correspond to the canonical disk-like ($n=1$) and bulge-like ($n=4$) \sersic values.
      The discretized values along the dust attenuation axis correspond to the sampling of this parameter used for our \texttt{CIGALE} fits (see Sect.~\ref{subsec:cigale}).
      }
      \label{fig:sersic}
   \end{figure*}
From the full structural parameter analysis performed on our $S/N>75$ sample, we can study for the first time the rest-MIR \sersic indices of 35 mass-complete SFGs at $0.1<z<2.5$.
The distribution of these rest-MIR \sersic indices is shown in the left panel of Fig.~\ref{fig:sersic distri} and is compared to that observed in the rest-optical. 

The rest-MIR light profiles of our SFGs are mostly consistent with exponential light profiles, with a median $n_{\rm MIR}$ value of $0.7_{-0.3}^{+0.8}$ (here and hereafter, the range corresponds to the 16th and 84th percentiles). 
None of our galaxies have a rest-MIR \sersic index compatible with a bulge-like light profile (i.e., $n\gtrsim3$), although our simulations has shown that if they exist, we should be able to measure them (Sect.~\ref{sec: MCMC}).
The rest-MIR axis ratio distribution (i.e., $b/a = 1- \varepsilon$; Fig.~\ref{fig: Ellipticity}) of our SFGs follows that of local spirals \citep[e.g.,][]{Rodriguez.2013}.
Together with their exponential light profiles, this demonstrates that these SFGs have a disk-like morphology in the rest-MIR. 
We cannot, however, rule out the presence of a sub-dominate bulge-like star-forming core in these galaxies, because with the angular resolution and $S/N$ of these MIRI observations, we are unable to perform robust double-component \sersic fits.

The rest-MIR \sersic index distribution resembles that observed in the rest-optical, although the latter is shifted toward slightly higher values with a median $n_{\rm Opt.}$ of $0.9_{-0.3}^{+1.2}$.
However, a two-sample Kolmogorov-Smirnov (KS) analysis cannot rule out that these two samples are drawn from the same distribution ($p$-value$\,=0.20$).
While these rest-MIR and rest-optical \sersic index distributions agree as a population, this agreement is reduced when considering each galaxy individually (right panel of Fig.~\ref{fig:sersic distri}). 
Indeed, the relation between $n_{\rm MIR}$ and $n_{\rm Opt.}$ is characterized by a mild correlation and a relatively large dispersion.
This suggests that while both the stellar and dust-obscured star-forming components of these galaxies have a disk-like morphology, their exact spatial distributions are intrinsically different.

We then studied the possible correlation between the rest-MIR \sersic indices of galaxies and some of their key physical properties such as their redshift, stellar mass, rest-MIR size, axis ratio, dust attenuation ($A_{V}$), and distance to the MS (Fig.~\ref{fig:sersic}).
We find no clear correlation, although the small number statistics offered by our $S/N>75$ sample may hide subtle relations to some extent.

\begin{figure}
   \centering
   \includegraphics[angle=0,width=\linewidth]{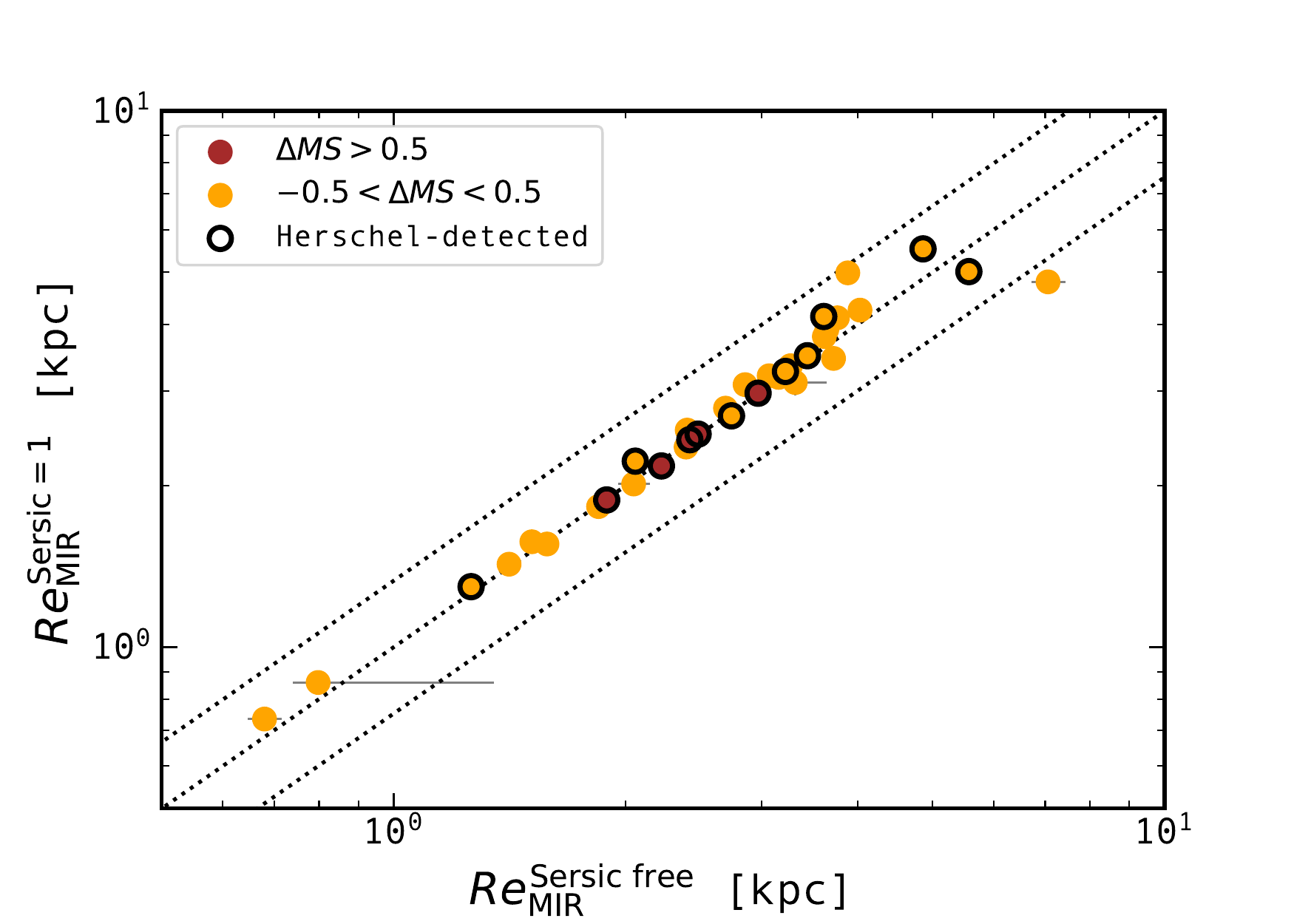}
      \caption{Comparison between the effective radius inferred by setting or not $n_{\rm MIR}$ to a value of 1 when fitting the 35 SFGs with $S/N>75$ in our final sample.
      Circles are color-coded by the distance of each galaxy to the MS, i.e., $-0.5 < \Delta MS < 0.5$ (orange), and $\Delta MS > 0.5$ (brown).
      Circles outlined by black edges are detected in the FIR by \textit{Herschel}.
      The dotted lines show the one-to-one relation and relative errors of $\pm33\%$.
      }
      \label{fig:fixing sersic}
   \end{figure}

Finally, we tested on this sample the impact of fixing $n_{\rm MIR}$ to a value of one during our partial structural parameter analysis.
In Fig.~\ref{fig:fixing sersic}, we compare the effective radius inferred for these 35 galaxies using our full and partial structural parameter analysis.
Fixing $n_{\rm MIR}$ to a value of one does not seem to introduce any significant and systematic bias into our partial structural parameter analysis. 
The inferred radii ratio has a mean value of 0.98 and a dispersion of 0.10.
As already mentioned, this moderate impact is due to the fact that our SFGs intrinsically have \sersic indices very consistent with a value of one and that $\Delta Re/\Delta n$ remains low when $n\sim1$ and $\Delta n <1$.
Nevertheless, this systematic error is not accounted for in our statistical error measurements.
We thus scaled these statistical errors by a factor of 1.1 in our partial structural parameter analysis.

\subsection{Rest-MIR sizes \label{subsec: sizes}}
By running our partial structural parameter analysis on all $S/N>10$ galaxies, we now extend our study to our mass-complete sample of 69 SFGs at $0.1<z<2.5$.
Although the \sersic indices of these galaxies are fixed to a value of one, we can recover their effective radii and ellipticities, as we have shown in Sect.~\ref{sec: MCMC} and \ref{subsec: sersic}.  
\subsubsection{Size--mass relations}
\begin{figure*}
   \centering
   \includegraphics[angle=0,width=\linewidth]{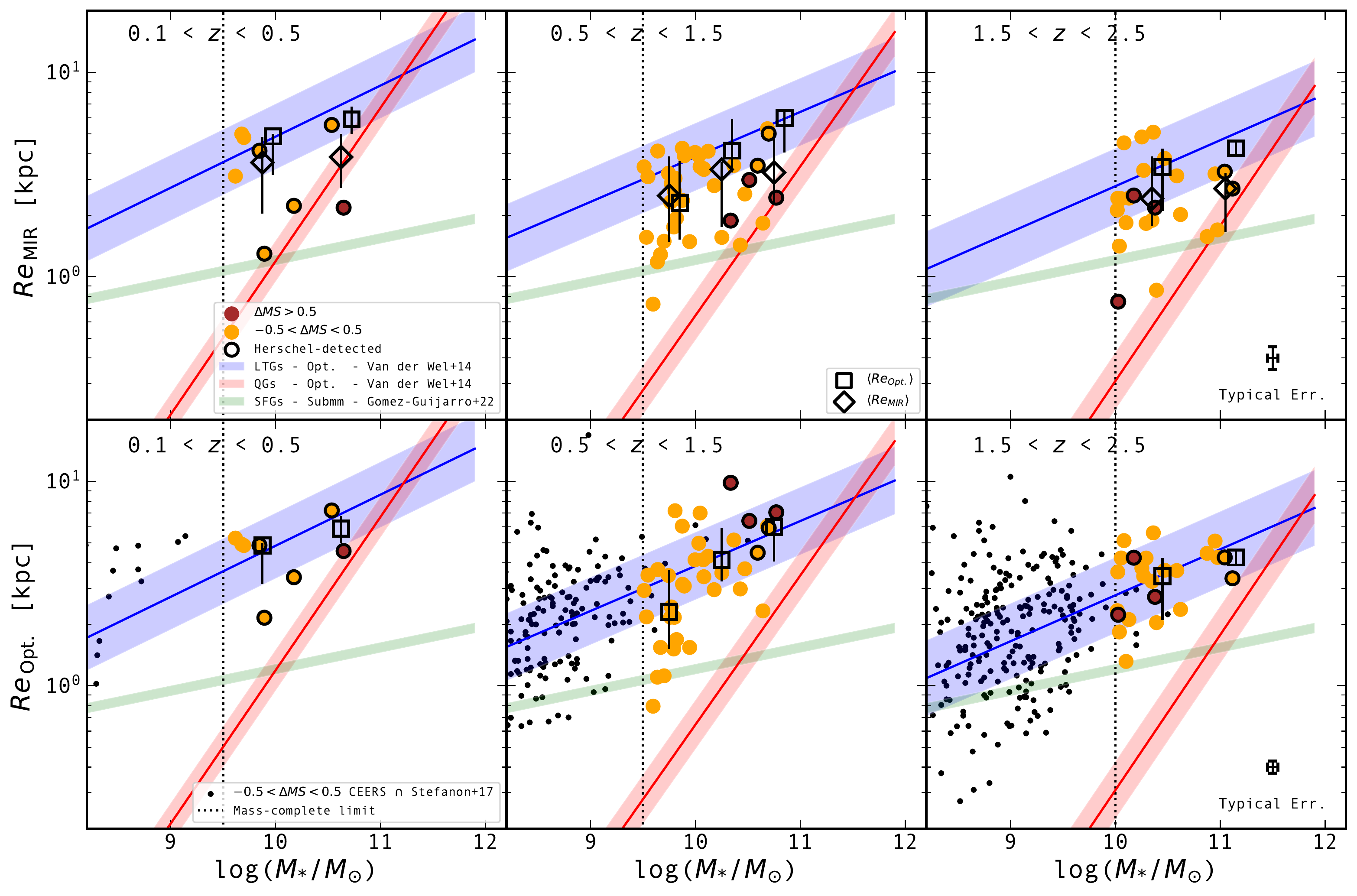}
      \caption{Rest-MIR size--mass (\textit{upper panels}) and rest-optical size--mass (\textit{lower panels}) distributions of the 69 SFGs in our final sample.
      Circles are color-coded by the distance of each galaxy to the MS, i.e., $-0.5 < \Delta MS < 0.5$ (orange), and $\Delta MS > 0.5$ (brown).
      Circles outlined by black edges are detected in the FIR by \textit{Herschel}.
      The rest-MIR sizes were inferred by setting $n_{\rm MIR}=1$ during our partial structural parameter analysis.
      Black dots show galaxies within the four CEERS red MIRI fields but which remained undetected in a MIRI bandpass that is dominated by dust emission (i.e., $S/N<10$ or $S_{\nu}^{\rm dust}/S_{\nu}^{\rm total} <75\%$).
      Open diamonds and squares show the 16th, 50th, and 84th percentiles of the rest-MIR and rest-optical size distributions in bins of stellar mass, respectively. 
      Squares (rest-optical sizes) are reproduced in the upper panels for comparison to the rest-MIR sizes and have been shifted by 0.1\,dex along the $x$-axis for clarity.
      The blue- and red-shaded regions correspond to the size--mass relation inferred in the rest-optical by \citet{Wel.2014} for late-type and early-type galaxies, respectively.
      The green-shaded region shows the size--mass relation inferred in the submm by \citet{Gomez-Guijarro.2022} using mostly massive ($\sim10^{10.8}\,M_{\odot}$) and high-redshift ($z\sim2.5$) SFGs. 
      Here, we assumed a mean axis ratio of 0.5 (see Fig.~\ref{fig: Ellipticity}) to convert their circularized radii into semi-major axis radii.
      Vertical dotted lines show the mass completeness limits of our sample.
      Typical $1\sigma$ error bars for individual objects are shown in the upper and lower right panels.
      }
      \label{fig:size--mass}
   \end{figure*}

One of the most fundamental relations involving the effective radius of galaxies is the so-called size--mass relation \citep[e.g.,][]{Wel.2014}.
In the upper panels of Fig.~\ref{fig:size--mass}, we present the rest-MIR size vs. mass distribution in our sample and compare it to the rest-optical size--mass relation inferred by \citet{Wel.2014}.
To first order, our rest-MIR size vs. mass distribution is characterized by an increase of the rest-MIR sizes with stellar mass. 
However, at high stellar masses, our galaxies appear to have smaller rest-MIR sizes than those predicted by the rest-optical size--mass relation, and some of our galaxies even have rest-MIR sizes that are as small as those predicted for quiescent galaxies. 
Because this finding could, nevertheless, be due to a combination of small number statistics and a relatively large dispersion in the rest-optical size--mass relation, we also show in the lower panels of Fig.~\ref{fig:size--mass}, the rest-optical size vs. mass distribution in our sample. 
In doing so, we unambiguously reveal that while for low stellar masses ($\lesssim10^{10.5}\,M_{\odot}$), the rest-MIR and rest-optical sizes are consistent within the uncertainties, for high stellar masses, the rest-MIR sizes are significantly smaller than those at rest-optical wavelengths (see open squares and diamonds).
This finding is qualitatively and quantitatively in line with the results of \citet{Shen.2023} who used a preliminary CEERS MIRI dataset on a UV-selected sample of SFGs.
Although this finding is also in qualitative agreement with recent ALMA studies that found compact rest-submm sizes relative to the rest-optical size--mass relation \citep[e.g.,][]{Lang.2019, Puglisi.2019, Tadaki.2020, Puglisi.2021,Gomez-Guijarro.2022}, our rest-MIR sizes are larger than the dust-obscured star formation sizes inferred from rest-submm \citep[e.g., $\sim$$\,$$2.8\,$kpc vs. $\sim\,$$1.4\,$kpc at $\sim\,$$10^{10.5}\,M_\odot$ and $z$$\,\sim\,$$2$;][]{Gomez-Guijarro.2022}. 
We note, however, that this latter study is dominated by massive ($\sim10^{10.8}\,M_{\odot}$) and high-redshift ($z\sim2.5$) galaxies and thus comparison with the MIRI sample mostly relies on extrapolation of their rest-submm relation to lower masses and redshifts.
The comparison between our results at rest-MIR and those at rest-submm with ALMA is discussed in depth in Sect.~\ref{subsec: compact galaxies}.

\subsubsection{Rest-optical to rest-MIR size ratio}
\begin{figure*}
   \centering
   \includegraphics[angle=0,width=\linewidth]{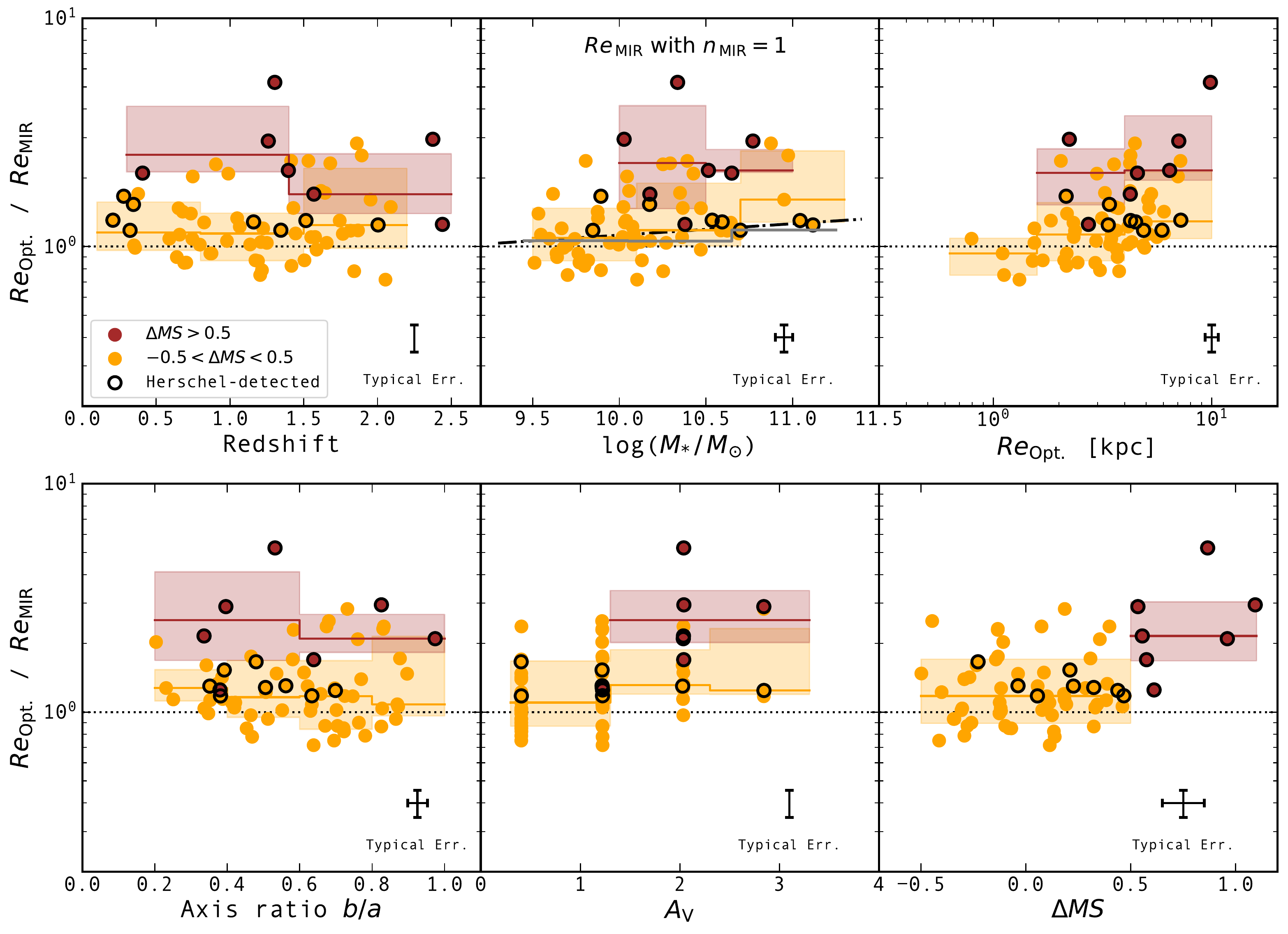}
      \caption{Rest-optical to rest-MIR size ratios for the 69 SFGs in our final sample as a function of their redshifts, stellar masses, rest-optical sizes, axis ratio, dust attenuation, and distances to the MS.
      Circles are color-coded by the distance of each galaxy to the MS, i.e., $-0.5 < \Delta MS < 0.5$ (orange), and $\Delta MS > 0.5$ (brown).
      Circles outlined by black edges are detected in the FIR by \textit{Herschel}.
      Brown line and shaded region show the median and 16th and 84th percentiles of starbursts ($\Delta MS > 0.5$) in bins of redshift, stellar mass, rest-optical effective radius, axis ratio, dust attenuation, and distance to the MS, respectively.
      Orange regions display the same quantities but for MS galaxies.
      These rest-MIR sizes were inferred by setting $n_{\rm MIR}=1$ during our partial structural parameter analysis.
      In the upper central panel, we show as dash-dotted line the mass-dependent rest-optical to rest-NIR size ratio found in \citet{Suess.2022}, and as gray line, the median for MS galaxies when replacing the rest-optical sizes of 18 galaxies by their rest-NIR sizes measured on their MIRI $7.7\,\mu$m images which are, according to our \texttt{CIGALE} SED fits, dominated by their stellar component.
      Typical $1\sigma$ error bars for individual objects are shown in each panel.
      The discretized values along the dust attenuation axis correspond to the sampling of this parameter used for our \texttt{CIGALE} fits.
      }
      \label{fig:size ratio}
   \end{figure*}

In order to analyze in more detail this mass-dependent relation between the rest-MIR and rest-optical sizes, we study in Fig.~\ref{fig:size ratio} the dependence of the rest-optical to rest-MIR size ratio as a function of some key physical properties of our galaxies, that is, their redshifts, stellar masses, rest-optical sizes, axis ratio, dust attenuation ($A_{V}$), and distances to the MS.
It is important to note that we do not find a significant absolute offset between the rest-optical and rest-MIR centers of these galaxies, with a median and standard deviation astrometric offset of 0.02\arcsec (160\,pc at $z=1$) and 0.04\arcsec (320\,pc at $z=1$), respectively.

\begin{figure}
   \centering
   \includegraphics[angle=0,width=\linewidth]{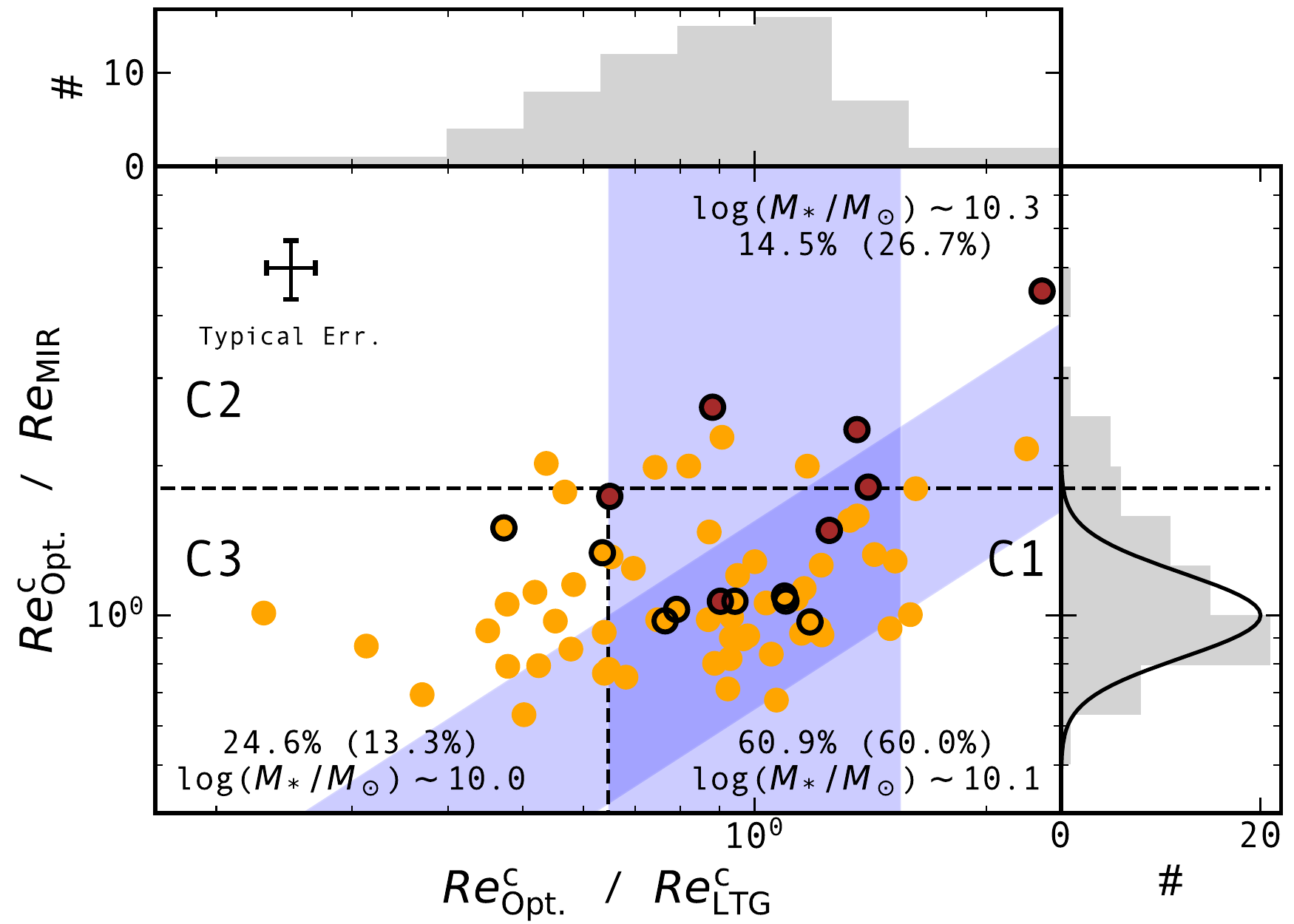}
      \caption{Rest-optical to rest-MIR size ratios for the 69 SFGs in our final sample as a function of the distance of their rest-optical sizes to the size--mass relation of LTGs (i.e., $Re^{\rm c}_{\rm Opt.}/Re^{\rm c}_{\rm LTG}(M_\ast,z)$).
      Right and upper panels present the distributions of these two quantities.
      Circles are color-coded by the distance of each galaxy to the MS, i.e., $-0.5 < \Delta MS < 0.5$ (orange), and $\Delta MS > 0.5$ (brown).
      Circles outlined by black edges are detected in the FIR by \textit{Herschel}.
      A typical $1\sigma$ error bar for individual objects is shown in the upper left.
      Galaxies within the vertical (diagonal) blue-shaded region have their rest-optical (rest-MIR) sizes well within the size--mass relation of LTGs \citep[$\sigma\sim0.19\,$dex;][]{Wel.2014}.
      The numbers in each frame provide the median stellar mass as well as the fraction of galaxies in each of these three populations above our mass completeness limits and above $>10^{10.5}\,M_{\odot}$ in parenthesis: galaxies with extended star-forming and stellar components (C1); galaxies with a compact star-forming component embedded in a extended stellar component (C2); and finally, galaxies with compact star-forming and stellar components (C3). 
      } 
      \label{fig:LTG distances}
   \end{figure}

Four main conclusions can be drawn from Fig.~\ref{fig:size ratio}.
Firstly, galaxies above the MS (i.e., starbursts) have rest-MIR sizes that are, independently of their redshifts and stellar masses, a factor $\sim2$ smaller than their rest-optical sizes.
This is the most significant trend observed in Fig.~\ref{fig:size ratio}.
Starbursts are frequently associated with major mergers \citep[e.g.,][]{Hung.2013,Cibinel.2019,Pearson.2019} and indeed the \textit{HST}-F160W Gini and $M_{20}$ coefficients of our starbursts, measured using the python package \texttt{Statmorph}, are shifted, compared to our MS galaxies, toward the merger region defined by \citet{Lotz.2008}.
Our finding that rest-MIR sizes are $\sim2$ times smaller than rest-optical sizes supports thus the notion that merger-driven starbursts develop dusty, star-forming cores amidst disturbed optical morphologies; as observed, for example, in the local antennae galaxies \citep[see also][]{Puglisi.2019}.
Secondly, the median rest-optical to rest-MIR size ratio of MS galaxies increases with their stellar masses, from $1.1_{-0.2}^{+0.4}$ at $\sim10^{9.8}\,M_{\odot}$ to $1.6_{-0.3}^{+1.0}$ at $\sim10^{11}\,M_{\odot}$.
Thirdly, the distribution of rest-optical to rest-MIR size ratios of MS galaxies is not Gaussian but rather skewed towards high values ($\gtrsim1.8$; see also the right panel of Fig.~\ref{fig:LTG distances}). 
This implies that there is a population of galaxies with compact star-forming components relative to their stellar components.
Fourthly, the rest-optical to rest-MIR size ratio does not evolve with redshift and does not depend on the axis ratio.
   
Although the finding of increasingly larger rest-optical to rest-MIR size ratios could be interpreted as, for example, the signature of the formation of dense stellar cores in massive galaxies, one should be cautious before drawing such a conclusion.
Indeed, these rest-optical sizes (at rest-frame $5000\,\AA$) could still be affected by negative radial color gradients and be different from the intrinsic half-stellar mass sizes of these galaxies \citep[e.g.,][]{Suess.2019, Suess.2022, Miller.2023,Zhang.2023}; this effect being probably more important in massive SFGs, which have larger dust attenuation than lower mass systems \citep[][]{Pannella.2015, Gomez-Guijarro.2023,Zhang.2023}.
Larger color gradients at higher stellar masses would thus explain the increase of the rest-optical to rest-MIR size ratio with stellar mass and dust attenuation seen in Fig.~\ref{fig:size ratio}.
To investigate the effect of radial color gradient on stellar size measurements, \citet{Suess.2022} used the CEERS \textit{JWST} F150W and F440W images and compared the rest-optical sizes of $\sim\,$1,000 SFGs at $1.0<z<2.5$ with their rest-NIR sizes (at rest-frame $1.5\,\mu$m), the latter being considered the best possible proxy for the stellar mass distribution in these galaxies.
They find a mass-dependent rest-optical to rest-NIR size ratio, suggesting that large radial color gradients in massive galaxies affect their rest-optical sizes, which therefore differ from their intrinsic half-stellar mass sizes.
We show in the upper central panel of Fig.~\ref{fig:size ratio}, the mass-dependent rest-optical to rest-NIR size ratio found in \citet{Suess.2022}. 
This relation strikingly matches that observed in our sample between our rest-optical and rest-MIR sizes.
It comes that at all stellar masses, $Re_{\rm Opt.}/Re_{\rm MIR} \sim Re_{\rm Opt.}/Re_{\rm NIR}$, and thus $Re_{\rm NIR}/Re_{\rm MIR} \sim 1$.
This implies that the mass-dependent rest-optical to rest-MIR size ratio observed in our study is likely due to radial color gradients affecting the rest-optical size measurements and that the median rest-MIR sizes are consistent, at all stellar masses, with the rest-NIR sizes and therefore with the half-stellar mass sizes of these galaxies.

Eighteen of our galaxies have a MIRI $7.7\,\mu$m band detection which is, according to our \texttt{CIGALE} SED fit (Sect.~\ref{subsec:cigale}), dominated by their stellar component (i.e., $S_{\nu}^{\rm stellar}/S_{\nu}^{\rm total} >75\%$; e.g., ID 12091; see upper panel of Fig.~\ref{fig:cigale}).
Replacing the rest-optical sizes of these 18 galaxies by their rest-NIR sizes measured on the MIRI $7.7\,\mu$m image, significantly reduces the mass-dependent trend observed in the upper central panel of Fig.~\ref{fig:size ratio} (see gray line in this panel).
This reinforces the idea that our rest-optical sizes are affected by radial color gradients at high masses and that the extents of the star-forming and stellar components of the SFGs probed by our study are, on average, consistent.
Future \textit{JWST} surveys with wide overlapping NIRCam and MIRI coverage will be essential to confirm these findings by allowing detailed measurement of the shape of the radial attenuation profiles of SFGs.

We then investigate the evolution of the rest-MIR-to-rest-optical size ratios with the distance of the rest-optical sizes from the size--mass relation of SFGs (i.e., late-type galaxies, LTGs; $Re_{\rm Opt.}/Re_{\rm LTG}(M_\ast,z)$; Fig.~\ref{fig:LTG distances}).
Because we found that our rest-optical sizes and the rest-optical size--mass relation of \citet{Wel.2014} were likely affected by color gradients, we corrected them using the rest-NIR-to-rest-optical size ratio versus mass relation of \citet{Suess.2022}, effectively turning these rest-optical sizes into half-mass sizes.
In Fig.~\ref{fig:LTG distances}, we identify three populations of galaxies.
Firstly, the largest population (C1), which contains about $61\%$ of our galaxies and corresponds to systems with an extended stellar component (i.e., sitting within or above the size--mass relation when considering their rest-optical sizes; $Re^{\rm c}_{\rm Opt.}/Re^{\rm c}_{\rm LTG}(M_\ast,z) > 10^{-0.19}$, i.e., within 1$\sigma$ of the size--mass relation) and an equally extended star-forming component (i.e., $Re^{\rm c}_{\rm Opt.}/Re_{\rm MIR} < 1.8$).
This $Re^{\rm c}_{\rm Opt.}/Re_{\rm MIR} \sim 1.8$ limit was set to three time the dispersion of a Gaussian function fit to the distribution of rest-optical to rest-MIR size ratios with $Re^{\rm c}_{\rm Opt.}/Re_{\rm MIR} < 1$ (black line in Fig.~\ref{fig:LTG distances}).
Secondly, a population (C3) that contains about $24\%$ of our galaxies and corresponds to systems with a compact stellar component ($Re^{\rm c}_{\rm Opt.} / Re^{\rm c}_{\rm LTG}(M_\ast,z) < 10^{-0.19}$) and an equally compact star-forming component (i.e., $Re^{\rm c}_{\rm Opt.}/Re_{\rm MIR} < 1.8$).    
This population of galaxies has structural properties close to those of the so-called blue nuggets, that is, galaxies which have already built their dense stellar core and could be on their way to quiescence \citep[e.g.,][]{Barro.2017}.
Thirdly, a population (C2) that contains about $15\%$ of our galaxies and corresponds to systems with a compact dusty star-forming component embedded in a larger stellar structure (i.e., $Re^{\rm c}_{\rm Opt.}/Re_{\rm MIR} > 1.8$). 
Most of these galaxies also have a stellar component that lies well within the size--mass relation (i.e., $Re^{\rm c}_{\rm Opt.}/Re^{\rm c}_{\rm LTG}(M_\ast,z) > 10^{-0.19}$) and which can therefore be considered as extended.
This new population of galaxies, which cannot be identified solely using optical images, could be the missing link between the galaxies with extended stellar component sitting on the size--mass relation and the blue nuggets with their already compact stellar component (see discussion in Sect.~\ref{subsec: evol. path}).
We note that the fractions of galaxies in each of these three populations remain about the same if we exclude starbursts (i.e., $\Delta MS>0.5$) or focus only on the high masses (i.e., $M_\ast>10^{10.5}\,M_\odot$). 

\subsubsection{Evolution of the stellar density, $\Sigma_{1}$}

   \begin{figure}
   \centering
   \includegraphics[angle=0,width=\linewidth]{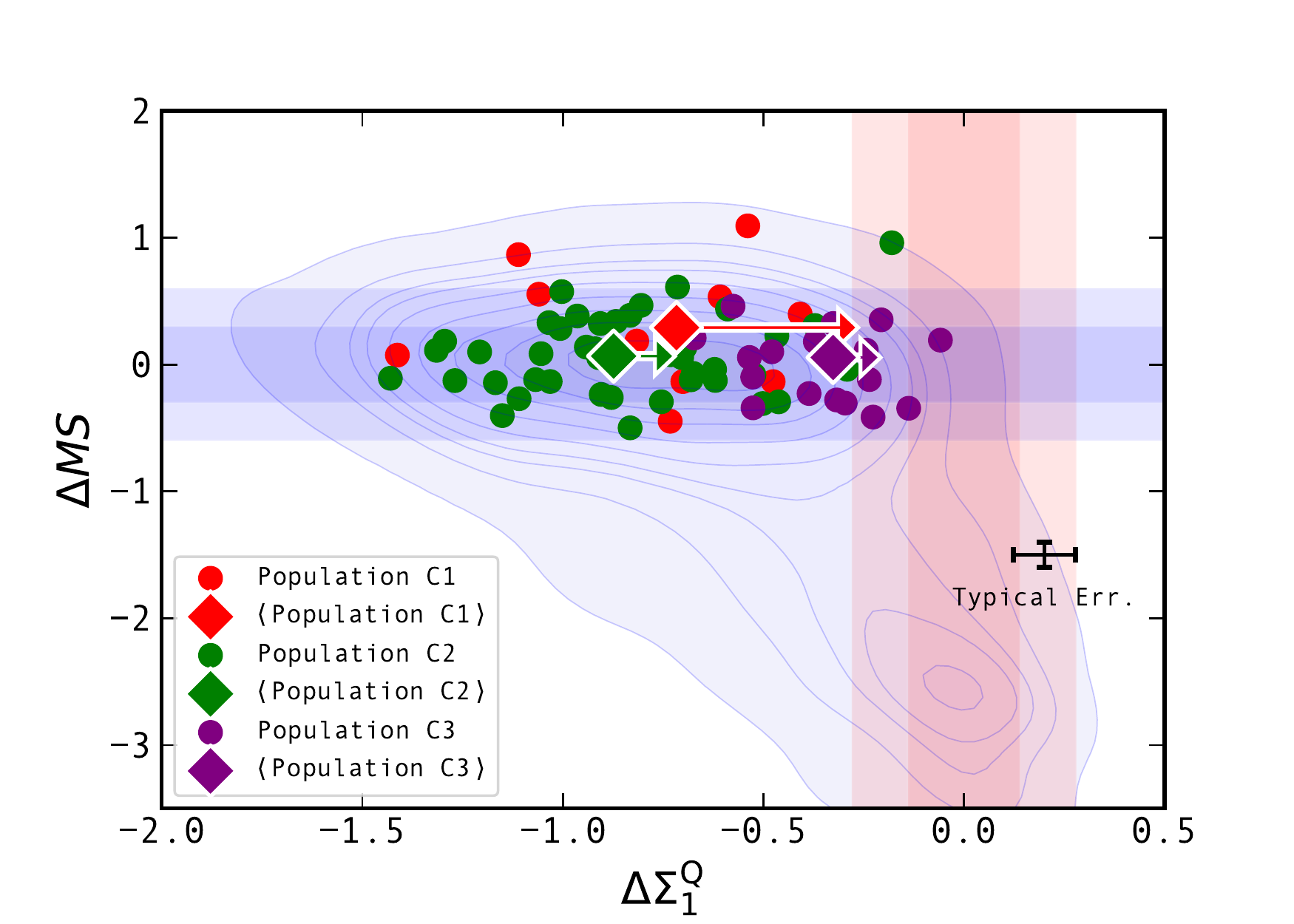}
      \caption{Distance to the MS as a function of the distance to the mass--$\Sigma_{1}$ relation of QGs \citep[$\Delta \Sigma_1^{\rm Q} \equiv {\rm log}\,(\Sigma_{1}^{\rm Q}(M_\ast,z)/\Sigma_1)$; see][]{Barro.2017} for the 69 SFGs in our final sample.
      Green circles correspond to the population of galaxies with extended star-forming and stellar components (C1); red circles correspond to the population of galaxies with a compact star-forming component embedded in an extended stellar component (C2); and finally, purple circles correspond to the population of galaxies with compact star-forming and stellar components (C3).
      Green, red, and purple diamonds show the median values of these populations.
      The median increases of $\Delta \Sigma_1^{\rm Q}$ for these three different populations of galaxies are displayed by right-pointing arrows (see text for details of this calculation).
      Contours correspond to the distribution of all sources in the EGS, plotted using a kernel density estimator. 
      The blue regions show the 1- and 2-$\sigma$ dispersion of the MS, while the red regions show the 1- and 2-$\sigma$ dispersion of the mass--$\Sigma_{1}^{\rm Q}$ relation.
      A typical $1\sigma$ error bar for individual objects is shown in the bottom right.
      }
      \label{fig:Delta Sigma 1}
   \end{figure}
The stellar density within the inner 1\,kpc radius of a galaxy (i.e., $\Sigma_{1}$) is known to be a key structural parameter for SFGs and QGs \citep[e.g.,][]{Barro.2017, Mosleh.2017, Gomez-Guijarro.2019, Suess.2021}.
For example, QGs have significantly higher core densities than SFGs ($\Sigma_{1}^{\rm Q}(M_\ast,z)> \Sigma_{1}^{\rm SFG}(M_\ast,z)$), which suggests that on their way to quiescence SFGs must experience a phase of significant core growth relative to the average evolution of their $\Sigma_{1}$ along the $\Sigma_{1}^{\rm SFG}$--mass relation \citep[e.g.,][]{Barro.2017}.
This is illustrated by the fact that the distribution of all galaxies in the EGS relative to the MS and to the $\Sigma_{1}^{\rm Q}$--mass relation follows an L-shape (see blue contours of Fig.~\ref{fig:Delta Sigma 1}), that is, SFGs within the MS but already exhibiting a dense stellar core (i.e., blue nuggets; $\Sigma_{1} \sim \Sigma_{1}^{\rm Q}(M_\ast,z)$ and thus $\Delta \Sigma_1^{\rm Q}\sim0$) could be the precursors of QGs emerging at later times.
We note that here $\Sigma_{1}$ has been calculated using the rest-optical effective radius and \sersic index of these galaxies. We also show in Fig.~\ref{fig:Delta Sigma 1} the distribution of our galaxies in the $\Delta MS-\Delta\Sigma_{1}$ plane \citep[using as proxies of their $\Sigma_{1}$ their rest-optical effective radius and \sersic index, correcting the former for color gradients using the rest-NIR-to-rest-optical size ratio versus mass relation of][]{Suess.2022}.
As expected, our galaxies are distributed over a broad range of $\Delta\Sigma_{1}$: from SFGs with relatively extended stellar component and thus light core densities to blue nuggets with their compact stellar component and thus dense core densities\footnote{Blue nuggets are formally defined in \citet{Barro.2017} as SFGs within 2-$\sigma$ of the $\Sigma_{1}^{\rm Q}$--mass relation.}.
Using the size of the star-forming component of these galaxies, we can now go one step further than previous studies and predict the evolution along the $\Delta\Sigma_{1}$ axis that the on-going star formation will produce.
To do so, we first assumed that the molecular gas reservoir of each of our galaxies is given by the scaling relation ($M_{\rm gas} = f(M_{\ast},z,\Delta MS)$) of \citet{Wang.2022kir}, extrapolated, when needed, to lower stellar masses ($<10^{10}\,M_\odot$) than those probed by their study.
Then, we simply measured the increase in $\Sigma_1$ assuming that all this molecular gas is turn into stars within 600\,Myrs \citep[i.e., the typical molecular gas depletion time of these galaxies;][]{Wang.2022kir} and that the newly formed stars is distributed following a \sersic index of one and according to the rest-MIR sizes of these galaxies.
In doing so, we find that galaxies with an already compact stellar component and equally compact star-forming component (C3) will continue to move closer to the $\Sigma_{1}^{\rm Q}$--mass relation, well into the blue nuggets region (i.e., within 2-$\sigma$ of the $\Sigma_{1}^{\rm Q}$--mass relation) and are thus likely the precursors of QGs.
In addition, we find that while the population of galaxies with extended stellar and star-forming components (C1) will also move toward the $\Sigma_{1}^{\rm Q}$--mass relation, their core densities will not approach those observed in blue nuggets.
On the contrary, we find that galaxies with an extended stellar component but a compact star-forming component (C2) will have, within the next 600\,Myrs, core densities consistent with those observed in blue nuggets.
This significant increase in stellar density is naturally mostly due to the fact that star formation in this population C2 is currently taking place in compact star-forming regions.

\section{Discussion \label{sec:discussion}}
We studied the stellar (i.e., rest-optical) and dust-obscured star formation (i.e., rest-MIR) morphologies (i.e., sizes and \sersic indices) of SFGs at $0.1<z<2.5$.
We find that (i) the rest-MIR \sersic index of bright galaxies ($S/N>75$) has a median value of $0.7_{-0.3}^{+0.8}$, which, together with their rest-MIR axis ratio distribution, suggests a disk-like morphology; (ii) the median rest-optical to rest-MIR size ratio of MS galaxies increases with their stellar mass, from $1.1_{-0.2}^{+0.4}$ at $\sim10^{9.8}\,M_{\odot}$ to $1.6_{-0.3}^{+1.0}$ at $\sim10^{11}\,M_{\odot}$; (iii) there exists a minor population of SFGs ($\sim15\%$) with a compact star-forming component embedded in an extended stellar structure (i.e., $Re^{\rm c}_{\rm Opt.} > 1.8 \times Re_{\rm MIR}$). 
Because the unobscured SFRs of our galaxies (i.e., SFR$_{\rm UV}$) are subdominant compared to their dust-obscured SFRs \citep[$\rm{SFR}_{\rm IR}/({\rm SFR}_{\rm UV}+{\rm SFR}_{\rm IR})\gtrsim0.7$ according to our \texttt{CIGALE} fits; see also][]{Whitaker.2017,Shen.2023}, in what follows, we consider the star-forming component of our galaxies as accurately traced by their dust-obscured star-forming component.

\subsection{A population of compact SFGs \label{subsec: compact galaxies}}
Our analysis reveals that about 15\% (10 galaxies) of our mass-complete sample ($10\%$ if we consider only MS galaxies; 6 galaxies) have a compact star-forming component embedded in a more extended stellar structure\footnote{This fraction would be at most $25\%$ (20 galaxies) if we consider that all AGNs above our stellar mass complete limits and excluded from our analysis belong to this population.} (i.e., with $Re^{\rm c}_{\rm Opt.}/Re_{\rm MIR} > 1.8$).
While this finding is consistent with those obtained in the (sub)mm with ALMA \citep[e.g.,][]{Simpson.2015, Ikarashi.2015, Hodge.2016, Fujimoto.2017, Gomez-Guijarro.2018, Elbaz.2018, Lang.2019, Rujopakarn.2019, Puglisi.2019, Franco.2020, Chang.2020, Chen.2020, Tadaki.2020, Gomez-Guijarro.2022, Zavala.2022}, the fraction of SFGs with such compact star-forming component appears to be very different if one considers their rest-MIR or rest-submm emissions.

To date the largest ALMA studies on the dust continuum sizes of SFGs are those of \citet{Tadaki.2020} and \citet{Gomez-Guijarro.2022}.
\citet{Tadaki.2020} studied a mass–selected sample of 85 massive ($>10^{11}\,M_{\odot}$) SFGs at $z=1.9-2.6$, while \citet{Gomez-Guijarro.2022} focused on a flux-limited sample of 88 massive ($M_\ast^{\rm med}\sim10^{10.8}\,M_\odot$) SFGs at $z=1.5-4$.
Both studies find that almost all SFGs have smaller dust emission regions relative to the stellar structure, with a median optical-to-(sub)mm size ratio of $2-3$.
This implies that more than 50\% of their SFGs would satisfy our definition of a compact star-forming component embedded in an extended stellar structure.
Such population, three time more numerous than ours (i.e., 50\% vs 15\%), suggests that compact SFGs dominate at the higher masses and redshifts probed by the \citet{Tadaki.2020} and \citet{Gomez-Guijarro.2022} samples, and/or that the rest-MIR or the rest-submm emissions do not accurately trace the intrinsic sizes of the star-forming component of SFGs.

A fairer comparison in terms of redshift and stellar mass, is obtained by looking at the results of \citet{Puglisi.2021}.
In this study, they measured the effective radius of the CO line emission of a \textit{Herschel}-selected sample of 77 SFGs at $z=1.1-1.7$ and with $M_\ast = 10^{10} - 10^{11.5}\,M_\odot$.
They find that at least $46\%$ of  their SFGs exhibit a compact star-forming component embedded in an extended stellar structure.
Nevertheless, their definition of compact SFGs differs from ours as it is solely based on the distance of their rest-submm sizes from the size--mass relation of LTGs, that is, $Re_{\rm submm} < 0.5\times Re_{\rm Opt.}^{\rm LTG}(M_\ast,z)$.
A non negligible fraction of their compacts SFGs have, however, rest-NIR sizes that are also below the size--mass relation of LTGs.
Applying to their sample our more restrictive $Re^{\rm c}_{\rm Opt.}/Re_{\rm MIR} > 1.8$ criterion would lower their fraction of compact SFGs \citep[see Fig.~3 of][]{Puglisi.2019}, rendering their and our results in better agreement. 
In addition, because their sample is \textit{Herschel}-selected, it is only probing the MS at $>10^{10.5}\,M_\odot$ and it is biased toward starbursts at $<10^{10.5}\,M_\odot$.
If the fraction of compact SFGs increases with stellar mass and distance to the MS (this latter assumption is supported by our results), this could  explain the remaining disagreement between our rest-MIR and their rest-submm results.

To investigate in more detail the compatibility between our rest-MIR results and those obtained in the rest-submm with ALMA, rest-submm size measurements of a mass-selected sample of intermediate-mass MS galaxies at $z=1-3$ are needed.
Such constraints are, however, difficult to obtain as the detection at high angular resolution of the dust emission of intermediate-mass MS galaxies is observationally very expensive even with ALMA. 
To alleviate this problem, \citet{Wang.2022kir} measured through stacking the mean rest-submm sizes of a mass-complete sample of MS galaxies at $z=0.5-3$ and with $M_\ast>10^{10}\,M_\odot$.
With this statistical approach, they find that the mean rest-submm size of MS galaxies does not evolve significantly with redshift nor stellar mass, with a mean circularized half-light radius of 2.2 kpc.
Assuming a mean axis ratio of 0.5 (see Fig.~\ref{fig: Ellipticity}), this translates into a mean effective semi-major axis radius of 3.1\,kpc, in line with our rest-MIR results (with a median value of $\sim3\,$kpc at $z>0.5$ and $M_\ast>10^{10}\,M_\odot$; see Fig.~\ref{fig:size--mass}).
While this agreement does not tell us anything about the fraction of compact SFGs as a function of stellar mass and redshift, it suggests that on average rest-MIR and rest-submm emission are probing roughly the same star-forming size.
Naturally, we cannot rule out that on the galaxy-to-galaxy basis, large radial dust temperature gradients affecting the rest-submm-to-SFR relation \citep{Popping.2021ve} or large radial metallicity and radiation hardness gradients affecting the MIR-to-SFR relation \citep{Schreiber.2017q5r}, would yield different rest-MIR and rest-submm size measurements.
   \begin{figure*}
   \centering
   \includegraphics[angle=0,width=\linewidth]{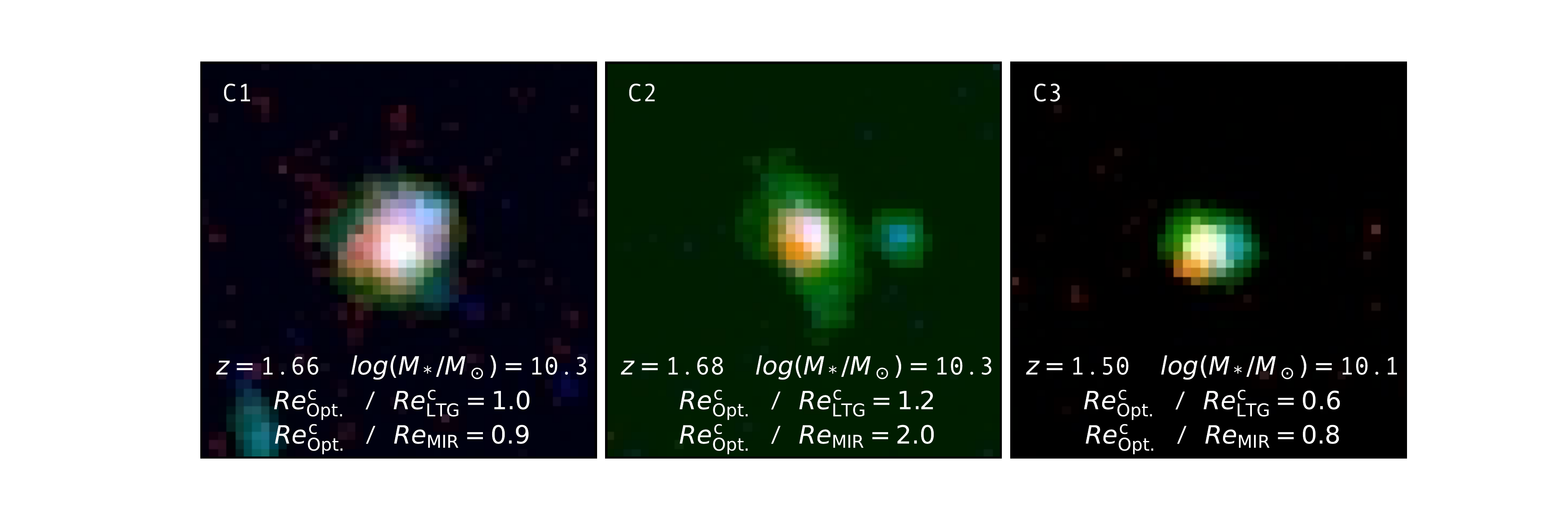}
      \caption{Composite images (R: \textit{JWST} F1500W -- G: \textit{HST} F160W -- B: \textit{HST} F606W) of a representative of the C1 (galaxies with extended star-forming and stellar components), C2 (galaxies with a compact star-forming component embedded in a extended stellar component), and C3 (galaxies with compact star-forming and stellar components) populations defined in Fig.~\ref{fig:LTG distances}. Each image has a size of $5\arcsec\times5\arcsec$. 
      }
      \label{fig:C1 C2 C3 RGB}
   \end{figure*}
   
Although some discrepancies remain, our results in the rest-MIR and those in the rest-submm, unambiguously reveal that a non-negligible fraction ($15\%$) of MS galaxies have compact star-forming component embedded in a more extended stellar structure.
The origin of these compact SFGs is still highly debated.
Based on the gas content and starburst-like excitation properties \citep{Puglisi.2021} of these galaxies, \citet{Gomez-Guijarro.2022xk} propose two scenarios to explain their formation. 
A first scenario in which a gas-rich merger funnels gas to the center of collision and enhances star formation in a compact star-forming region.
As a result, the galaxy moves well above the MS.
After few hundreds Myrs, as the gas reservoir is consumed, the SFR declines and the galaxy crosses, incidentally, the MS still exhibit its compact star-forming component. 
A second scenario in which angular momentum loss, driven externally (mergers or counter-rotating streams) or internally (clump migration), funnels gas to the center of the galaxy, enhances star formation in a moderately compact star-forming region, although not well above the MS; as the gas reservoir is consumed the star-forming region becomes more compact (outside-in gas consumption) and more star formation efficient, sustaining thereby the galaxies on the MS for a relatively long time.
In the case of the first scenario, the fraction of compact SFGs at a given stellar mass is thus given by the product of the mass function of SFGs and the fraction of starbursts at a given stellar mass.
Because the fraction of starbursts is roughly constant \citep[5\%;][]{Schreiber.2015} and the stellar mass function of SFGs has a very steep shape \citep[e.g.,][]{Weaver.2022}, one would expect that in the case of the first scenario the fraction of compact SFGs in the MS should remain constant or even decrease with mass.
In contrast, in the second scenario the fraction of compact SFGs in the MS is supposed to increase with mass.
A significant increase with mass of the fraction of compact SFGs is suggested by the comparison of our results with those of \citet{Tadaki.2020} and \citet{Gomez-Guijarro.2022xk}, which favors the second scenario. 

\subsection{From extended SFGs to compact SFGs to blue nuggets, a possible evolutionary sequence \label{subsec: evol. path}}
The rest-MIR sizes of MS galaxies appear to be on average smaller than those at rest-optical wavelengths and this trend increases with the stellar mass; $Re_{\rm Opt.} /Re_{\rm MIR} \sim 1.1_{-0.2}^{+0.4}$ at $\sim10^{9.8}\,M_{\odot}$ to $1.6_{-0.3}^{+1.0}$ at $\sim10^{11}\,M_{\odot}$.
This could be interpreted as a sign that the cold gas accreted by MS galaxies is efficiently channelled into their central region and triggers the formation of their bulges \citep[e.g.,][]{Tonini.2016}. 
Instead, we argue that this mass-dependent trend is mostly due to un-corrected negative radial color gradients, which increases their apparent rest-optical sizes; an effect that increases with stellar masses. 
The existence of discrepant half-light and half-mass radii in the rest-optical is supported both by observations \citep{Suess.2019, Lang.2019, Suess.2022}, simulations \citep{Popping.2021ve,Costantin.2023}, and radiative transfer modeling of galaxies' observed dust and structural properties \citep{Zhang.2023}.
Correcting statistically for this effect using the mass-dependent relation of \citet{Suess.2022}, reveals that in most SFGs (85\%), the star-forming and stellar components have the same size (see Fig.~\ref{fig:LTG distances}).
This population of galaxies with similar stellar and star-forming sizes must be, however, separated further into two different populations: a population of galaxies ($61\%$) sitting on the size--mass relation of LTGs and which have thus extended stellar and star-forming components; and a smaller population of galaxies ($24\%$) falling below the size--mass relation of LTGs and which have thus compact stellar and star-forming components (i.e., with structural properties close to those of the so-called blue nuggets).
Finally, a third population, representing the remaining $15\%$ of the SFG population, has a compact star-forming component embedded in a more extended stellar structure (i.e., with $Re^{\rm c}_{\rm Opt.}/Re_{\rm MIR} > 1.8$).
As an example, composite images of a representative of each of these populations (selected to have similar redshift and stellar mass) are shown in Fig.~\ref{fig:C1 C2 C3 RGB} (R: \textit{JWST} F1500W -- G: \textit{HST} F160W -- B: \textit{HST} F606W).

Qualitatively, these three populations of galaxies, namely, the main population of extended SFGs (C1), the compact SFGs discussed in Sect.~\ref{subsec: compact galaxies} (C2), and the population of galaxies with compact stellar component (C3), might describe an evolutionary sequence (C1$\rightarrow$C2$\rightarrow$C3) that follows the main phases of structural evolution in the $\Delta MS-\Delta\Sigma_{1}$ plane \citep[see][]{Barro.2017}: a nearly horizontal branch of MS disks growing inside-out (C1) that will transition into the compaction knee of the $\Delta MS-\Delta\Sigma_{1}$ plane (C3, i.e., $\sim$blue nuggets) as a result of a significant increase in central density (C2), and lastly these blue nuggets will quench after reaching a maximum central density.
In other words, the newly discovered population of MS galaxies with a compact star-forming component embedded in a more extended stellar structure would be the missing link between the main population of SFGs sitting on the size--mass and $\Sigma_{1}^{\rm SFG}$--mass relations and blue nuggets with their compact stellar component and which are on their way to quiescence.

Several observational evidences support this evolutionary sequence, although care must be taken when using galaxies in a limited redshift range to sketch an evolutionary pattern.
First, our main population of galaxies with extended stellar and star-forming components is qualitatively consistent with an inside-out growth scenario for disk formation \citep[e.g.,][]{Nelson.2016,Matharu.2022}.
Indeed, at least half of this population experiences ongoing star formation activity at preferentially larger radii than existing stars, and this fraction would further increase when taking into account their dust-unobscured star formation component, which is slightly larger on average than their stellar and dust-obscured star-forming components \citep{Shen.2023}.
The inside-out growth of this galaxy population is also suggested by the fact that their star-forming components have shallower profiles than their stellar component (i.e., $n_{\rm MIR}<n_{\rm Opt.}$), and thus that ongoing star formation would actually increase the size of their stellar component.
Then, as expected in this evolutionary sequence, the distributions of extended and compact SFGs in the $\Delta MS-\Delta\Sigma_{1}$ plane are very similar (see green and red data points in Fig.~\ref{fig:Delta Sigma 1}).
Finally, ongoing star formation in compact SFGs will increase their central densities to those observed in blue nuggets (see red arrow in Fig.~\ref{fig:Delta Sigma 1}).

Although compact SFGs seem to be the missing link between the extended SFG population and blue nuggets, the mechanism triggering a compact star-forming component embedded in a more extended stellar structure remains unknown (see Sect.~\ref{subsec: compact galaxies}). 
This phase may be associated with the compaction of SFGs predicted in hydrodynamical simulations \citep{Ceverino.2015, Zolotov.2015, Tacchella.2016,Lapiner.2023}, and which is typically associated with very dissipative processes (i.e., gas-rich) like major and minor mergers (see, e.g., Fig.~\ref{fig:C1 C2 C3 RGB}), interaction-driven gravitational instabilities, and/or strong disk instabilities \citep{Dekel.2009, Ceverino.2010, Dekel.2014u6d,Lapiner.2023}.
Future high-resolution observations of the stellar component (\textit{JWST}-NIRCam and -NIRSpec) and gas reservoir  (ALMA and NOEMA) of these compact SFGs will put strong observational constraints on the compaction mechanism, which is expected to take place before quenching \citep[e.g.,][]{Barro.2017}.

\section{Summary\label{sec:summary}}%
We combined \textit{HST} images from CANDELS with \textit{JWST} images from CEERS to measure the stellar and dust-obscured star formation distributions of 69 SFGs. 
The rest-optical structural parameters were inferred from the \textit{HST} F125W and F160W images, while those at rest-MIR were measured on the sharpest MIRI images (i.e., shortest wavelength) dominated by dust emission ($S_{\nu}^{\rm dust}/S_{\nu}^{\rm total} >75\%$), as inferred for each galaxy from our optical-to-FIR SED fits with \texttt{CIGALE}.
We restricted our rest-MIR full structural parameter analysis (i.e., size, \sersic index, and ellipticity) to our brightest MIRI sample (i.e., $S/N>\,75$; 35 galaxies), while we extended our study to fainter sources (i.e., $S/N>10$; 69 galaxies) by performing a partial structural parameter analysis (i.e., size and ellipticity) by setting their \sersic index to unity.
The extension to fainter galaxies allowed us to measure and compare the rest-MIR and rest-optical sizes of a mass complete sample ($>80\%$) of SFGs down to $10^{9.5}, 10^{9.5},$ and $10^{10}\,M_{\odot}$ at $z\sim0.3$, 1.0, and 2.0, respectively.
Because the unobscured SFRs of these galaxies (i.e., SFR$_{\rm UV}$) are subdominant compared to their dust-obscured SFRs (with $\rm{SFR}_{\rm IR}/({\rm SFR}_{\rm UV}+{\rm SFR}_{\rm IR})\gtrsim0.7$ according to our \texttt{CIGALE} SED fits), one can consider that the star-forming component of our galaxies is accurately traced by their dust-obscured star-forming component.
With this unique dataset, we find the following:
   \begin{enumerate}
      \item Bright galaxies ($S/N>75$) have rest-MIR \sersic indices with a median value of $n_{\rm MIR}=0.7_{-0.3}^{+0.8}$ (the range corresponds to the 16th and 84th percentiles) and the distribution of their rest-MIR axis ratio follows that of local spirals. The star-forming component of SFGs thus has a disk-like morphology.
      \item The light profiles in the rest-MIR are, on average, slightly shallower than those in the rest-optical, whose median \sersic index is $n_{\rm Opt.}=0.9_{-0.4}^{+1.2}$ (although a two-sample KS analysis cannot rule out the possibility that these rest-optical and rest-MIR \sersic index distributions are drawn from the same distribution). There is only a mild correlation between $n_{\rm MIR}$ and $n_{\rm Opt.}$. Hence, while the stellar and star-forming components of SFGs both have a disk-like morphology, their exact spatial distributions are intrinsically slightly different.
      \item Galaxies above the MS (i.e., starbursts) have rest-MIR sizes that are, on average, a factor $\sim2$ smaller than their rest-optical sizes. Because starbursts are likely triggered by major mergers, these rest-optical sizes are probably dominated by the disrupted morphology of the merging stellar components, while the rest-MIR sizes are presumably dominated by the coalescing star-forming core.
      \item The median rest-optical to rest-MIR size ratio of MS galaxies increases with their stellar masses, from $1.1_{-0.2}^{+0.4}$ at $\sim10^{9.8}\,M_{\odot}$ to $1.6_{-0.3}^{+1.0}$ at $\sim10^{11}\,M_{\odot}$. This mass-dependent trend resembles that found in \citet{Suess.2022} between the rest-optical and rest-NIR sizes of SFGs, which suggests that it is mostly due to radial color gradients affecting the rest-optical sizes. There is no significant offset between the rest-optical and rest-MIR centers of these galaxies, with a absolute median and standard deviation astrometric offset of 0.02\arcsec and 0.04\arcsec, respectively.
      \item Correcting statistically for the effect of radial color gradients using the mass-dependent relation of \citet{Suess.2022}, reveals that the median rest-optical to rest-MIR size ratio of our SFGs is $1.0_{-0.2}^{+0.6}$. In most SFGs ($85\%$), the star-forming and stellar components thus have the same size. Of these galaxies, most fall within the size--mass relation of LTGs ($61\%$ of our total SFG sample) and thus have extended stellar and star-forming components; while a smaller fraction ($24\%$ of our total SFG sample) falls below the size--mass relation of LTGs and thus have compact stellar and star-forming components, that is, structural properties close to those of the so-called blue nuggets.
      \item There exists a minor population of SFGs ($\sim15\%$) with a compact star-forming component embedded in a more extended stellar component (i.e., $Re^{\rm c}_{\rm Opt.} > 1.8 \times Re_{\rm MIR}$). The ongoing star formation in these compact SFGs will be sufficient to increase their central densities (i.e., $\Sigma_1$; bulge formation) to those observed in blue nuggets.
   \end{enumerate}
The three populations of galaxies revealed by our study could describe an evolutionary sequence: from disk-like SFGs growing inside-out (i.e., the extended SFG phase) that will transition into the compaction knee of the $\Delta MS-\Delta\Sigma_{1}$ plane (i.e., the blue nugget phase) as a result of a significant increase in their central density (i.e., the compact SFG phase).
The compact SFGs would thus be the missing link between the main population of extended SFGs and the blue nuggets, the latter being on the way to quiescence.
The mechanism responsible for triggering such compact star-forming components remains unknown but is expected to be associated with highly dissipative (i.e., gas-rich) processes such as major and minor mergers, interaction-driven gravitational instabilities, and/or strong disk instabilities \citep{Dekel.2009, Ceverino.2010, Dekel.2014u6d}.
Future high-resolution observations of the stellar (\textit{JWST}-NIRSpec) and gas reservoir (ALMA and NOEMA) dynamics of compact SFGs will provide strong observational constraints on the mechanisms responsible for the formation of such a compact star-forming component embedded in a larger stellar structure.

\begin{acknowledgements}
      We would like to thank the referee for their comments that have helped to improve our paper.
      This work is based on observations made with the NASA/ESA/CSA James Webb Space Telescope. These observations are associated with program \#1345.
      PGP-G acknowledges support  from  Spanish  Ministerio  de  Ciencia e Innovaci\'on MCIN/AEI/10.13039/501100011033 through grant PGC2018-093499-B-I00.
      This work was supported by UNAM-PAPIIT IA102023.
      BM, DE, and CGG acknowledges support from CNES.
      BM acknowledges the following open source software used in the analysis: \texttt{Astropy} \citep{Collaboration.2022}, \texttt{photutils} \citep{Bradley.2022}, \texttt{NumPy} \citep{Harris.2020}, and \texttt{Statmorph} \citep{Rodriguez-Gomez.2018}.
\end{acknowledgements}

\end{document}